\documentclass[superscriptaddress,aps,pra,twocolumn,showpacs,nofootinbib,longbibliography,showkeys]{revtex4-2}
\usepackage{slashbox}
\usepackage{amsmath,amssymb,amsthm}
\usepackage{easybmat,comment}
\usepackage[colorlinks=true,citecolor=blue,urlcolor=blue, linkcolor = magenta]{hyperref}
\usepackage[pdftex]{graphicx}
\usepackage{times,txfonts}
\usepackage{braket}
\usepackage{color}
\usepackage{natbib}
\usepackage{subcaption}
\usepackage{ragged2e}
\DeclareCaptionJustification{justified}{\justifying}
 \usepackage[compat=0.4]{yquant}
\newcommand{\be}{\begin{equation}}
	\newcommand{\ee}{\end{equation}}
\newcommand{\ba}{\begin{eqnarray}}
	\newcommand{\ea}{\end{eqnarray}}

\newtheorem{theorem}{Theorem}

\usepackage{comment}
\usepackage{tikz}
\usetikzlibrary{quantikz}
\usepackage{graphicx}
\DeclareMathOperator\erf{erf}
\begin{document}
	
%\preprint{APS/123-QED}
\title{Entanglement-assisted continuous-variable concatenated codes for encoding qubits or oscillators}
%qubit (or oscillator)-into-oscillators }
\author{Nihar Ranjan Dash\textsuperscript{}}
   \email{dash.1@iitj.ac.in}
   \affiliation{Indian Institute of Technology Jodhpur, Rajasthan 342030, India}
\author{Sanjoy Dutta\textsuperscript{}}
    \email{sanjoy@ppisr.res.in}
    \affiliation{Poornaprajna Institute of Scientific Research (PPISR), Bidalur post, Devanahalli, Bengaluru 562164, India}
\author{R. Srikanth\textsuperscript{}}
   \email{srik@ppisr.res.in}
   \affiliation{Poornaprajna Institute of Scientific Research (PPISR), Bidalur post, Devanahalli, Bengaluru 562164, India}
\author{Subhashish Banerjee}
   \email{subhashish@iitj.ac.in}
   \affiliation{Indian Institute of Technology Jodhpur, Rajasthan 342030, India}

	%\textsuperscript{b,c}}
%\affiliation{
	%Third institution, the second for Charlie Author
%\affiliation{%


%\date{}% It is always \today, today,

\begin{abstract}
Entanglement-assisted (EA) stabilizer codes enhance the rate of error correction in relation to codes with no pre-shared entanglement. Meanwhile, bosonic error-correcting codes, such as the Gottesman-Kitaev-Preskill (GKP) code, can be concatenated with qubit stabilizer codes to significantly reduce the logical failure probability of those stabilizer codes. First, we combine the above two concepts to propose an EA version of the qubit-into-oscillators concatenated code that chains an EA-stabilizer (outer) code with a GKP (inner) code. As an example we present a three-qubit EA-repetition concatenated with a GKP code. Second, we propose an EA version of the non-Gaussian oscillator-into-oscillators concatenated code that chains a GKP (outer) code with an EA-stabilizer (inner) code. As an example we present a GKP code concatenated with a three-qubit EA repetition code that uses two maximally entangled modes (emodes) and suppresses the variances of both position and momentum quadrature errors of a data mode. Furthermore, we generalize the latter example to a family of GKP code concatenated with a $n$-qubit EA repetition code that uses {$n-1$} emodes and suppresses the variances of both position and momentum quadrature errors of a data mode by a factor ${1/n}$.
\end{abstract}
%\keywords{Keywords..}
%Use showkeys class option if keyword
\maketitle

%\tableofcontents

\section{Introduction}
Quantum stabilizer codes \cite{gottesman1997stabilizer, lidar2013quantum}, which form an essential approach for combating errors, involve encoding a smaller number of logical qubits into a larger number of physical qubits. Stabilizer codes have been studied extensively for both discrete-variable (DV) \cite{Laflamme1996perfect, shor1995scheme, steane1996error}, continuous-variable (CV) systems \cite{Braunstein1998error, lloyd1998analog} as well as for hybrid DV-CV systems \cite{lee2013near, kapit2016hardware, liu2026hybrid}, such as qubit-oscillator codes. Such hybrid codes concatenate CV (often bosonic) error-correcting codes with stabilizer qubit (DV) codes in order to improve the error-correcting performance. With recent advancements in the architecture of hardware processor systems, these hybrid codes offer a powerful computational prototype by integrating the strengths of both DV and CV systems \cite{liu2026hybrid}. Particularly, the Gottesman-Kitaev-Preskill (GKP) code \cite{gottesman2001encoding, grimsmo2021quantum} is one of the most promising bosonic error-correcting codes, suitable to correct small shift errors in the position and momentum quadrature of phase space. 

The GKP code has been concatenated as the inner layer with several DV outer codes such as the repetition code \cite{fukui2017analog, wang2019quantumerrorcorrectionGKP}, the five-qubit code \cite{lin2025exploring},  Knill's four-qubit and six qubit codes \cite{fukui2017analog, fukui2018tracking}, the surface codes \cite{fukui2018highthreshold, vuillot2019quantum, noh2020faulttolerant, noh2022lowoverhead}, the quantum lower-density parity check code \cite{Raveendran2022finiterateqldpcGKP}, the quantum convolutional code \cite{xiao2022quantum}. Ref. \cite{fukui2017analog} has proposed the use of the analog information of GKP qubits to improve the error correction performance when concatenated to an outer, three-qubit repetition code. A similar system, but with a generalized $n$-qubit repetition code concatenated with GKP codes, has been studied with both ideal and finite-energy GKP ancillas under biased Gaussian shift errors \cite{li2024correcting}. Such conventional concatenations of an outer stabilizer code and inner GKP code (denoted $\mathcal{C}_{\rm stabilizer}\rhd\mathcal{C}_{\rm GKP}$ codes or, more broadly, qubit-into-oscillators codes) find applications in quantum repeaters \cite{rozpkedek2021quantum} and fault-tolerant architecture designs \cite{chamberland2022ftqc}.

In an alternative approach, a single oscillator data mode with infinite dimension can be encoded into multiple physical oscillator modes, resulting in an oscillator-into-oscillators code \cite{noh2020encoding}. The \textit{$\mathcal{C}_{\rm GKP}\rhd\mathcal{C}_{\rm stabilizer}$} code is a subclass of the oscillator-into-oscillators code that can correct shift errors in both position and momentum quadratures of the data mode. Examples of $\mathcal{C}_{\rm GKP}\rhd\mathcal{C}_{\rm stabilizer}$ codes would be a GKP outer code concatenated with a repetition inner code ($\mathcal{C}_{\rm GKP}\rhd\mathcal{C}_{\rm repetition}$) and GKP-two-mode-squeezing code \cite{noh2020encoding}. A further example is the GKP outer code concatenated with Steane's seven qubit inner code ($\mathcal{C}_{\rm GKP}\rhd \mathcal{C}_{\rm Steane}$) code architecture whose experimental feasibility in CV platforms has recently been studied \cite{guo2025concatenateddualdisplacementcode}. Ref. \cite{xu2023qubit-oscillator} proposes qudit analogues of $\mathcal{C}_{\rm GKP}\rhd\mathcal{C}_{\rm stabilizer}$ codes to protect DV computational GKP states. The same authors also define an unbiased GKP-repetition code without additional quadrature squeezing; these codes can suppress Gaussian random shift errors. The optimal encoding of the $\mathcal{C}_{\rm GKP}\rhd\mathcal{C}_{\rm stabilizer}$ codes and their reduction to the GKP-two-mode-squeezing code are derived in \cite{Wu2023optimalencodingof}. The practical importance of CV resources and concatenation is underscored by the fact that $\mathcal{C}_{\rm GKP}\rhd\mathcal{C}_{\rm stabilizer}$ codes find application in distributed quantum sensing protocols \cite{Zhuang_2020, zhou2022enhanching} and quantum communication \cite{Wu_2022_QST}.

Entanglement-assisted (EA)-stabilizer codes arise naturally in a communication scenario when the stabilizer group with the sender is relaxed to admit non-Abelian operators \cite{brun2006correcting}. EA-stabilizer codes can improve the code transmission rates at the cost of pre-shared entanglement between the sender and receiver \cite{brun2014catalytic}. EA-stabilizer codes for CV systems were first studied in \cite{wilde2007optics}, in particular, a CV (bosonic mode) version of the EA-stabilizer code (what we call CV $\rhd$ EA-stabilizer) that corrects arbitrary single-mode errors by using two-mode maximally entangled states.

In this work, we consider the problem of augmenting concatenated CV-stabilizer codes with entanglement assistance, specifically the concatenation of qubit EA-stabilizer codes and oscillator (GKP) codes. The resulting EA concatenated codes inherit the resource thrift of EA codes and the enhanced correction capability of concatenated codes. In this work, we use the following notation: the expression $A \rhd B$ denotes a concatenation where $A$ (resp., $B$) is the outer (resp., inner) code. We extend a recent work comparing $\mathcal{C}_{\rm stabilizer}\rhd\mathcal{C}_{\rm GKP}$ and $\mathcal{C}_{\rm GKP}\rhd\mathcal{C}_{\rm stabilizer}$ codes \cite{xu2023qubit-oscillator} to a study of their EA counterparts, $\mathcal{C}_{\rm EA-stabilizer}\rhd\mathcal{C}_{\rm GKP}$ and $\mathcal{C}_{\rm GKP}\rhd\mathcal{C}_{\rm EA-stabilizer}$ codes. \color{black} Additionally, we consider the problem of  optimizing the entanglement assistance to these concatenated CV-stabilizer codes in terms of ancilla consumption or logical failure probability, applicable both to encoding qubits or oscillators. Here we use the ancillas that are computational GKP (in the case of $\mathcal{C}_{\rm EA-stabilizer}\rhd\mathcal{C}_{\rm GKP}$) or canonical GKP states (in the case of $\mathcal{C}_{\rm GKP}\rhd\mathcal{C}_{\rm EA-stabilizer}$). In prior research on $\mathcal{C}_{\rm GKP}\rhd\mathcal{C}_{\rm stabilizer}$ codes, error suppression in a single quadrature utilizing the $\mathcal{C}_{\rm GKP}\rhd\mathcal{C}_{\rm repetition}$ code, or error suppression in both quadratures through additional quadrature squeezing, has been reported \cite{noh2020encoding}. A scheme for error suppression in both quadratures using a total of $2n+1$ modes is known \cite{xu2023qubit-oscillator}. Here, we present $\mathcal{C}_{\rm GKP}\rhd\mathcal{C}_{\rm EA-repetition}$ code that uses the Gaussian encoder of the $[[3,1,3;2]]$ EA-repetition code and can protect a single data mode from both position and momentum quadrature errors by utilising two maximally entangled modes (emodes). We generalize it to a family of GKP code concatenated with a $n$-qubit EA repetition code that can suppress both position and momentum quadrature errors of a data mode with a noise suppression factor ${1/n}$ by utilizing ${n-1}$ emodes. 

After presenting preliminaries on EA-stabilizer codes and GKP codes (Sec. \ref{sec:prelim}), we propose and study EA versions of the concatenation of qubit stabilizer codes and GKP code in Sec. \ref{sec:concatenated_CV_EA_stabilizer}. In specific, in Section \ref{sec:A}, we demonstrate an EA version of the qubit-into-oscillator concatenated code, namely $\mathcal{C}_{\rm EA-stabilizer}\rhd\mathcal{C}_{\rm GKP}$ codes, with an example of $\mathcal{C}_{[[3,1,3;2]]} \rhd \mathcal{C}_{\rm GKP}$. Here, we encode each physical qubit and halves of the maximally entangled qubits (ebits) of the outer EA-stabilizer code by a computational GKP code. Reversing the concatenation order, in Section \ref{sec:B}, we demonstrate an EA version of the oscillator-into-oscillators concatenated codes, namely the $\mathcal{C}_{\rm GKP}\rhd\mathcal{C}_{\rm EA-stabilizer}$ codes, discussing the specific example of $\mathcal{C}_{\rm GKP}\rhd\mathcal{C}_{[[3,1,3;2]]}$ code. In Section \ref{sec:GKP_rhd_n_EA_rep}, we generalize the study to a family of GKP code concatenated with an $n$-qubit EA repetition code, namely the $\mathcal{C}_{\rm GKP}\rhd\mathcal{C}_{[[n,1,n;n-1]]}$ code. Crucially, the encoder employed is not the standard oscillator-into-oscillators type \cite{noh2020encoding}, but instead that used for EA-stabilizer codes. Here we compute the noise suppression in both quadratures arising from the use of entanglement assistance. We compare our suppression results with related results in the corresponding codes without entanglement assistance. Finally, we summarize and conclude in Sec. \ref{sec:conclusions}.

\section{Preliminaries}\label{sec:prelim}
\subsection{EA-stabilizer codes}
The EA-stabilizer codes are a generalization of stabilizer codes to admit a non-abelian set of stabilizers, and for amenability of construction from classical error-correcting codes, irrespective of self-duality \cite{brun2006correcting, brun2014catalytic}. An $[[n,k,d;c]]$ EA-stabilizer code has distance $d$ and encodes $k$ information qubits into $n+c$ physical qubits, where Alice transmits $n$ qubits to Bob, and they share $c$ ebits. The encoded state is given by \cite{lidar2013quantum}
\begin{align}
    \ket{\psi_{L}}&=\mathcal{U}^{\rm EA}(\ket{\psi}) \nonumber\\ & \equiv (U_{\rm enc} \otimes I_{B})(\ket{\psi} \otimes \ket{0}^{\otimes (n-k-c)} \otimes (\ket{\Phi^{+}}_{AB}^{\otimes c})).
    \label{eq: logicalstateeaqecc}
\end{align}
The sender, Alice, uses a unitary encoder $U_{\rm enc}$ to encode the information state $\ket{\psi}$ in addition to the ancilla and halves of the ebits from the maximally entangled state $\ket{\Phi^{+}}_{AB}$ pre-shared between herself and the receiver. 

\subsection{GKP states}
Gottesman-Kitaev-Preskill (GKP) states are the discrete states in CV systems designed in analogy with shift-resistant codes in DV systems \cite{gottesman2001encoding, grimsmo2021quantum}. Here, the encoding of a qubit protects against shift errors in canonical variables. More generally, we can encode qubits in the oscillator's $n$ modes. Let $\hat{q}=(\hat{a}^{\dagger}+\hat{a})/\sqrt{2}$ and $\hat{p}=i(\hat{a}^{\dagger}-\hat{a})/\sqrt{2}$ be the position and momentum quadrature operators, respectively, of a bosonic mode, where $\hat{a}$ and $\hat{a}^{\dagger}$ are the annihilation and creation operators with commutation relation $[\hat{a},\hat{a}^{\dagger}]=1$. 
The computational GKP states are
\begin{align}
    \ket{j_{\rm GKP}}&=\sum_{s \in \mathbb{Z}} \ket{\hat{q}=(2s+j)\sqrt{\pi}},\quad (j=0,1),
    \label{eq:logical_GKP}
\end{align}
which are the $\pm1$ eigenstates of the logical Pauli operator $\hat{Z}_{\rm GKP}\equiv \exp{[i\sqrt{\pi}\hat{q}]}$.
Also, the complementary codewords are $\ket{+_{\rm GKP}} \equiv \frac{1}{\sqrt2}(\ket{0_{\rm GKP}} + \ket{1_{\rm GKP}}) = \sum_{s \in \mathbb{Z}} \ket{\hat{p}=2s\sqrt{\pi}}$ and $\ket{-_{\rm GKP}} \equiv \frac{1}{\sqrt2}(\ket{0_{\rm GKP}} - \ket{1_{\rm GKP}}) = \sum_{s \in \mathbb{Z}} \ket{\hat{p}=(2s+1)\sqrt{\pi}}$, which are, respectively, the $\pm1$ logical Pauli operators of the GKP code given by
$\hat{X}_{\rm GKP} \equiv \exp{[-i\sqrt{\pi}\hat{p}]}$.

Two of the stabilizers for the GKP code are:
\begin{align}
    \hat{S}_{q}&=\hat{Z}_{\rm GKP}^{2}=\exp{[i2\sqrt{\pi}\hat{q}]},\nonumber\\
    \hat{S}_{p}&=\hat{X}_{\rm GKP}^{2}=\exp{[-i2\sqrt{\pi}\hat{p}]}.
    \label{eq:stabilizers_GKP}
\end{align}
Specifically pertinent to our objectives are the
 gates $\hat{\rm F}_{\rm GKP}$ and ${\rm SUM}_{j \rightarrow k}$: \begin{align}
    \hat{\rm F}_{\rm GKP}&=\exp{[i\frac{\pi}{2}\hat{a}^{\dagger}\hat{a}]}\equiv \exp[i \frac{\pi}{4}(\hat{p}^2+\hat{q}^2)],\nonumber\\
     {\rm SUM}_{j \rightarrow k}&= \exp{[-i\hat{q}_{j}\hat{p}_{k}]},
\end{align}
which are, respectively, the logical Hadamard (Fourier) and CNOT operations. Under the action of the ${\rm SUM}_{j \rightarrow k}$ gate, the quadrature operators transforms as
\begin{align}
    (\hat{q}_j, \hat{q}_k) &\rightarrow (\hat{q}_j, \hat{q}_j + \hat{q}_k ),\nonumber\\
    (\hat{p}_j,\hat{p}_k) &\rightarrow (\hat{p}_j - \hat{p}_k,\hat{p}_k) .
\end{align}

Assume that independent and identically distributed (iid) Gaussian random shift errors on the $n$ modes. The noise $\mathcal{N}^{k}[0,\sigma]$ occurring on the $k$-th mode of the oscillator adds random errors in the position and momentum quadratures of the $k$-th mode, $\zeta_{q}^{(k)}$ and $\zeta_{p}^{(k)}$, respectively, and is described by a Gaussian distribution with zero mean and variance $\sigma^{2}$. For the correction to be successful, the absolute shift in a single physical GKP qubit must be less than $\frac{\sqrt{\pi}}{2}$, which is half the grid spacing. Thus the probability of successfully identifying a bit value is \cite{menicucci2014fault, fukui2017analog}
 \begin{align}
     p_{\rm succ}= \int_{-\sqrt{\pi}/2}^{\sqrt{\pi}/2} \frac{1}{\sqrt{2\pi \sigma^{2}}} \exp{\left(\frac{-\zeta^{2}}{2\sigma^{2}}\right)}\,d\zeta.
     \label{eq:physical_success_rate_GKP}
 \end{align}
Suppose the GKP encoded state $
 \ket{\psi_{\rm GKP}}=\alpha \ket{0_{\rm GKP}}+\beta \ket{1_{\rm GKP}}$
undergoes a position shift error $E_q(\zeta) \equiv e^{-i\zeta \hat{p}}$ (where, for correctability, the condition $|\zeta| < \frac{\sqrt{\pi}}{2}$ needs to be satisfied). Then the erroneous state of the data qubit is given by
$
 e^{-i\zeta \hat{p}}\ket{\psi_{\rm GKP}} = \alpha \sum_{s \in \mathbb{Z}} \ket{\hat{q}=2s\sqrt{\pi}+\zeta}+\beta \sum_{s \in \mathbb{Z}} \ket{\hat{q}=(2s+1)\sqrt{\pi}+\zeta} \equiv \ket{\psi_{\rm GKP}+\zeta}.
$ The position measurement can be indicated schematically as follows:

\begin{align}
 &\sum_{m \in \mathbb{Z}} \ket{\hat{q}=m\sqrt{\pi}+\zeta}_{D}\otimes \sum_{m \in \mathbb{Z}} \ket{\hat{q} = n\sqrt{\pi}}_A\nonumber\\&\xrightarrow[]{\rm SUM_{D\rightarrow A}}\sum_{s \in \mathbb{Z}} \ket{\hat{q}=m\sqrt{\pi}+\zeta}_{D}\otimes \sum_{s \in \mathbb{Z}} \ket{\hat{q}=(m+n)\sqrt{\pi}+\zeta}_{A},
\end{align}
where $D$ and $A$ denote the data and ancillary GKP states, respectively. The value of the shift error $\zeta$ is obtained by a destructive CV measurement on the ancilla, and subsequently corrected by applying the negative shift $-{\zeta}$ in the data qubit. This GKP error-correction scheme is known as the Steane scheme \cite{steane1996error}. The second scheme, known as the Knill-Glancy scheme \cite{glancy2006error, schmidt2022quantum}, replaces the SUM gate with a beam splitter and a squeezer. As an alternative approach, Knill's teleportation-based GKP error correction \cite{Knill2005scalable} uses a logical GKP Bell state as an ancilla and couples it to a data qubit via a beam splitter. The quadrature measurements of the beam splitter output are used for GKP error correction. 

\begin{figure}[htp]
\begin{subfigure}[t]{0.3\textwidth}
\centering
\begin{tikzpicture}
\node[scale=0.8] {
    \begin{quantikz}
\lstick{$\ket{0_{\rm GKP}}$}  &\gate{\hat{F}_{\rm GKP}} &\ctrl{1} &\qw \rstick[2]{$\ket{\Phi^{+}_{\rm GKP}}$}\\ 
\lstick{$\ket{0_{\rm GKP}}$}  &\qw &\gate{\rm SUM} &\qw 
\\ 
\end{quantikz}
};
\end{tikzpicture}
\caption{}
\label{subfig:encoding_circuit_GKP_Bell}
\end{subfigure}
\begin{subfigure}[t]{0.3\textwidth}
\centering
\begin{tikzpicture}
\node[scale=0.8] {
    \begin{quantikz}
\lstick{$\ket{{\rm GKP}}$}  &\qw&\gate[2]{{\textsc{BS}}}  &\qw &\qw\rstick[2]{$\ket{\Phi^{+}_{\rm GKP}}$}\\ 
\lstick{$\ket{{\rm GKP}}$}    &\qw&\qw &\qw&\qw
\\ 
\end{quantikz}
};
\end{tikzpicture}
\caption{}
\label{subfig:GKP_Bell_from_cano_gkp}
\end{subfigure}
 \caption{ Encoding circuits to realize the GKP Bell pairs via two methods: (a) using computational GKP basis, by application of the Fourier and SUM gates on two modes initialized with $\ket{0_{\rm GKP}}$. (b) using a two-mode canonical GKP state, sent through a $50:50$ beam-splitter (BS).}
 \label{fig:epr_GKP}
 \end{figure}
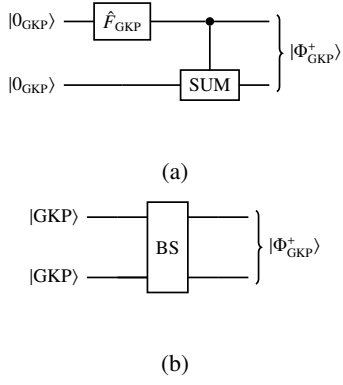

The measurement outcomes used to extract the syndromes 
\begin{align}
    z=R_{\sqrt{\pi}}(\zeta),
\end{align}
where $R_{\sqrt{\pi}}(\zeta) \equiv \zeta$ is the modular reduction funciton, if $\zeta\in [-\sqrt{\pi}/2,\sqrt{\pi}/2)$. Based on the measured syndromes, the estimated shift error $\tilde{\zeta}$ is calculated and subtracted from the actual quadrature error $\zeta$. This yields a residual logical error for the GKP qubit, given by
\begin{align}
    {\zeta}^{(*)}=\zeta-\tilde{\zeta}.
\end{align}

The canonical GKP state is a grid state with a narrower spacing (by factor $\sqrt{2}$) than the computational GKP state Eq. (\ref{eq:logical_GKP}) \cite{Wu2023optimalencodingof, xu2023qubit-oscillator}. Due to non-Gaussianity of the (canonical) GKP state, it is used in the encoding of the oscillator-into-oscillators codes to correct Gaussian errors, in accordance with the no-go result that asserts that correction of Gaussian errors can not be done by only Gaussian resources \cite{noh2020encoding}. The canonical GKP state, $\ket{\rm GKP}$ is given by
\begin{align}
    \ket{\rm GKP}&\propto \sum_{s \in \mathbb{Z}} \ket{\hat{q}=s\sqrt{2\pi}} \propto \sum_{s \in \mathbb{Z}} \ket{\hat{p}=s\sqrt{2\pi}}.
    \label{eq:canoGKP}
\end{align}
The GKP encoded Bell state $\ket{\Phi^{+}_{\rm GKP}}$ uses computational GKP states \cite{walshe2020continous}
\begin{align}
    \ket{\Phi^{+}_{\rm GKP}}=\frac{1}{\sqrt{2}}(\ket{0_{\rm GKP}}\otimes \ket{0_{\rm GKP}}+\ket{1_{\rm GKP}}\otimes \ket{1_{\rm GKP}}).
\end{align}
Analogous to the qubit Bell state, the GKP Bell state $\ket{\Phi^{+}_{\rm GKP}}_{AB}$ is stabilized by $\hat{Z}_{\rm GKP} \otimes \hat{Z}_{\rm GKP} =\exp[{i\sqrt{\pi}(\hat{q}_{A}+\hat{q}_{B}})]$ and $\hat{X}_{\rm GKP} \otimes \hat{X}_{\rm GKP} \equiv \exp[-{i\sqrt{\pi}(\hat{p}_{A}+\hat{p}_{B}})]$ \cite{schmidt2022quantum}. Measuring these stabilizers is equivalent to measuring $\hat{q}_{A}+\hat{q}_{B}$ and $\hat{p}_{A}+\hat{p}_{B}$ modulo $\sqrt{\pi}$. Two schemes for generating GKP Bell pairs are given in Fig. \ref{fig:epr_GKP}. Note that the use of finite-energy GKP states leads to asymmetric widening of one of the two modes of the finite-energy GKP Bell state in relation to the one produced by the first scheme \cite{marqversen2025performanceanalysisgkperror}.

\section{Concatenating CV and EA stabilizer codes}\label{sec:concatenated_CV_EA_stabilizer}

In general, concatenating two codes $[[n_1,k_1]]$ and $[[n_2,k_2]]$ yields two different resultant codes depending on the order of concatenation, i.e., $[[n_1,k_1]] \rhd [[n_2,k_2]] \ne [[n_2,k_2]] \rhd [[n_1,k_1]]$. It is straightforward to verify that if one of the codes is of the type $[[1,1]]$, then the resulting code is order-independent, i.e., $[[1,1]] \rhd [[n,k]] = [[n,k]] \rhd [[1,1]]$. (We can think of $[[1,1]]$ as a ``substitution code'' that substitutes a qubit of type I for one of type II.) Nevertheless the encoding and decoding procedures will be different in both cases. In the context of GKP states combined with stabilizer states, this leads to two corresponding types of encoding. 

Accordingly, one of them, what we call the $\mathcal{C}_{\rm stabilizer}\rhd\mathcal{C}_{\rm GKP}$  can be characterized as follows:
\begin{align}
    \ket{\psi_{\rm GKP}}&= (\mathcal{U}^{\rm GKP} \circ \mathcal{U}^{\rm stabilizer})\ket{\psi}. 
    \label{eq:stab_GKP_encoding}
\end{align}
Here the $[[n,k,d]]_2$ code word is first produced, and subsequently individual qubits are encoded into computational GKP states. During decoding, individual oscillators are corrected for CV errors and decoded into a $n$-qubit state, before discrete errors are corrected using the $[[n,k,d]]$ stabilizers, and then decoded. In the second scheme, what we call the $\mathcal{C}_{\rm GKP}\rhd\mathcal{C}_{\rm stabilizer}$ can be characterized as follows:
\begin{align}
    \ket{\psi_{\rm GKP}}&= (\mathcal{U}^{\rm stabilizer} \circ \mathcal{U}^{\rm GKP})\ket{\psi}. 
    \label{eq:GKP_stab_encoding}
\end{align}
Here a single-mode GKP code is first produced, and subsequently encoded into a multi-mode GKP code using the encoder of the stabilizer code. During decoding, modes are corrected for CV errors and decode into a single-mode GKP code. Note that though Eqs. (\ref{eq:stab_GKP_encoding}) and (\ref{eq:GKP_stab_encoding}) produce the same result, their resource requirements are different. When we replace the stabilizer code with an EA-stabilizer code, we correspondingly obtain two types of concatenated codes, discussed in the following two subsections.

\begin{figure}[t]
    \centering
    \includegraphics[width=6cm]{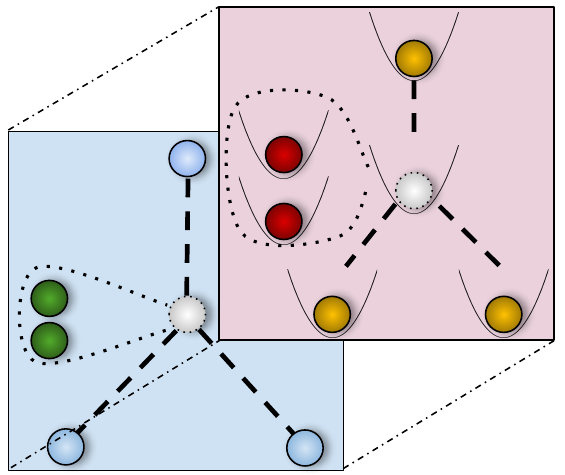}
    \caption{(Color online) code structure of an $\mathcal{C}_{[[3,1,3;2]]}\rhd\mathcal{C}_{\rm GKP}$ code with outer $[[3,1,3;2]]$ EA-repetition code and inner GKP code: A single logical qubit of the outer $[[3,1,3;2]]$ EA-repetition code represented as a white circle encoded into three physical qubits (blue circles) by utilizing two pre-shared entangled bits( green circles). The inner GKP code further encodes these three qubits, along with the two entangled bits, into harmonic oscillators, represented by yellow and red circles, respectively.}
     \label{fig:conc_code_str_ea_GKP}
\end{figure}

\subsection{EA-stabilizer $\rhd$ GKP codes \label{sec:A}}
We define $\mathcal{C}_{\rm EA-stabilizer}\rhd\mathcal{C}_{\rm GKP}$ codes as those encoded as follows:
\begin{align}
    \mathbb{Z}_2^{\otimes k} \xrightarrow{[[n,k,d;c]]_2}\mathbb{Z}_2^{\otimes (n+c)}\xrightarrow{\text{GKP}}\mathbb{R}^{\otimes (n+c)},
    \label{eq:ea_stab_GKP_encoding}
\end{align}
where $\mathbb{Z}_2$ and $\mathbb{R}$ are used to describe a qubit and oscillator system, respectively. Here the $[[n,k,d;c]]_2$ code word is first produced, and subsequently individual qubits along with ebits are encoded into computation GKP states. During decoding, individual modes along with part of the emodes are corrected for CV errors and decoded into a $n$-qubit state, before discrete errors are corrected using the $[[n,k,d;c]]$ EA stabilizers, and then decoded.

\begin{figure}
\centering
\begin{tikzpicture}
\node[scale=0.8] {
    \begin{quantikz}
\lstick{$\ket{\psi}_{1}$}  &\qw &\ctrl{3} &\gate{\rm CNOT} &\qw &\qw\\
& \makeebit[-60][]{$\ket{\Phi^{+}}_{2,4}$} &  \qw &\ctrl{-1} &\gate{\rm CNOT} &\qw\\
&     &\qw &\qw &\qw &\qw\\
& \makeebit[-60][]{$\ket{\Phi^{+}}_{3,5}$} &\gate{\rm CNOT} &\qw &\ctrl{-2} &\qw\\
&     &\qw &\qw &\qw &\qw
\end{quantikz}
};
\end{tikzpicture}
\caption{Encoding circuit for $[[3,1,3;2]]_2$ entanglement-assisted repetition code.}
\label{fig:encoding_circuit_3ea_rhd_GKP}
\end{figure}
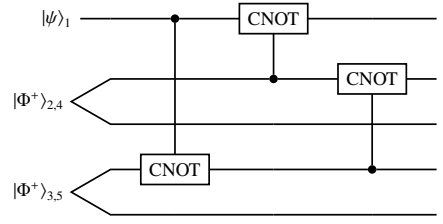

Codes here encode a GKP state into an outer EA-stabilizer code. We discuss a simple example to illustrate how the encoding, error correction and decoding works here.
We define the $\mathcal{C}_{\rm EA-repetition}\rhd\mathcal{C}_{\rm GKP}$ code as a class from the broad $\mathcal{C}_{\rm EA-stabilizer}\rhd\mathcal{C}_{\rm GKP}$ codes. Particularly, We consider the $[[3,1,3;2]]$ EA-repetition code from the family $[[n,1,n;n-1]]$ EA-repetition code with odd $n \geq 3$ \cite{lai2013entanglement}. First, the outer-layer encoding encodes the $3$ physical qubits into a single logical qubit using $2$ ebits. Next, the inner-layer encoding encodes each of the $3$ physical qubits, along with $2$ ebits, using GKP qubit-to-oscillator codes.  Here, we label the data mode as $1$, and the other two GKP emodes on Alice's side as $2$ and $3$. Encoding for the $\mathcal{C}_{[[3,1,3;2]]}\rhd\mathcal{C}_{\rm GKP}$ code is given by
\begin{align}
    \mathbb{Z}_2^{\otimes 1} \xrightarrow{[[3,1,3;2]]_2}\mathbb{Z}_2^{\otimes 5}\xrightarrow{\text{GKP}}\mathbb{R}^{\otimes 5}.
    \label{eq:31ea_stab_GKP_encoding}
\end{align}
The corresponding encoder is given by \cite{lai2014dualities}:
\begin{equation}
    U_{\rm enc} = \text{CNOT}_{3\rightarrow2}\;\text{CNOT}_{2\rightarrow1}\; \text{CNOT}_{1\rightarrow3},
    \label{eq:uenc}
\end{equation}
which takes as input the data qubit and two halves of ebits (Figure \ref{fig:encoding_circuit_3ea_rhd_GKP}).

After encoding, Alice sends the encoded state through a Gaussian random-shift channel, where physical GKP modes undergo shift errors that are subsequently corrected by applying negative shifts. 
Next comes the outer EA-stabilizer error correction. For the outer ${[[3,1,3;2]]}$ EA-repetition code, the logical operators $\hat{X}_{\rm GKP}$ and $\hat{Z}_{\rm GKP}$ are embedded with the stabilizers of the EA code. Thus, the form of the stabilizers required for the outer EA-stabilizer layer error correction in the $\mathcal{C}_{[[3,1,3;2]]}\rhd\mathcal{C}_{\rm GKP}$ code resulted as
\begin{align}
    \begin{matrix}
 \hat{X}_{\rm GKP} & \hat{X}_{\rm GKP} & \hat{I}  &\vline & \hat{X}_{\rm GKP} & \hat{I} \\
 \hat{I} & \hat{Z}_{\rm GKP} & \hat{Z}_{\rm GKP}  & \vline & \hat{Z}_{\rm GKP} & \hat{I} \\
 \hat{I} & \hat{X}_{\rm GKP} & \hat{X}_{\rm GKP}  & \vline & \hat{I} & \hat{X}_{\rm GKP} \\
 \hat{Z}_{\rm GKP} & \hat{Z}_{\rm GKP} & \hat{I}  & \vline & I & \hat{Z}_{\rm GKP}  \\
 \label{eq:stabilizers_3_1_3_2_rep_conc_GKP}
\end{matrix}.
\end{align}
Here, the bar indicates that the three modes ($1$, $2$, and $3$) are in Alice's possession, and modes $4$ and $5$ are in Bob's possession.

Consider the situation of concatenating two codes, $\mathcal{C}_{o}$ and $\mathcal{C}_{i}$, designating the outer and inner codes with logical failure probabilities $p_L^{(o)}(p)$ and $p_L^{(i)}(p)$, respectively. The concatenated scheme's logical failure probability is then \cite{gaitan2008quantum, dash2023concatenating}
\begin{align}
    p_L^{\rm CC}(p) = p_L^{(o)}[p_L^{(i)}(p)].
  \label{eq:logical_error_rate_cc}
\end{align}
 For the $[[3,1,3;2]]$ code, the logical failure probability is given by \cite{lai2012entanglement}
\begin{align}
  p_L^{[[3,1,3;2]]}& =1 - (1-p)^{3} - 3  p (1-p)^2\nonumber\\
  &- (2/9)  p^2  (1-p).
  \label{eq:logical_error_rate_3_1_3_2_rep}
\end{align}
The logical failure probability of the $\mathcal{C}_{[[3,1,3;2]]}\rhd\mathcal{C}_{\rm GKP}$ is obtained by replacing $p$ with $p_L^{\rm GKP}=1-p_{\rm succ}$ from Eq. (\ref{eq:physical_success_rate_GKP}) in Eq. (\ref{eq:logical_error_rate_3_1_3_2_rep}):
\begin{align}
    p_L^{\mathcal{C}_{[[3,1,3;2]]}\rhd\mathcal{C}_{\rm GKP}}&=p_L^{\rm [[3,1,3;2]]}[p_L^{\rm GKP}],\nonumber\\
    & = 1 - \frac{2}{9} \erf{\left(\frac{\sqrt{\pi}}{2\sqrt{2} \sigma}\right)}- \frac{23}{9} \erf{\left(\frac{\sqrt{\pi}}{2\sqrt{2} \sigma}\right)}^{2}\nonumber\\
    &+\frac{16}{9} \erf{\left(\frac{\sqrt{\pi}}{2\sqrt{2} \sigma}\right)}^{3},
    \label{eq:logical_error_rate_3_1_3_2_rep_GKP}
\end{align}
where $\text{erf}(x)$ is the error function. In Fig. \ref{fig:logical_error_rate_3eagkp_gkp3ea_5qubit}, we compare the logical failure probabilities of the $\mathcal{C}_{ [[3,1,3;2]]}\rhd\mathcal{C}_{\rm GKP}$ code and the $\mathcal{C}_{[[5,1,3]]}\rhd\mathcal{C}_{\rm GKP}$. The former outperforms the latter, which can be attributed to the fact that the outer code in the former has noiseless halves of the ebits in Bob's possession.

\subsection{GKP$\rhd$EA-stabilizer codes \label{sec:B}}
We define $\mathcal{C}_{\rm GKP}\rhd\mathcal{C}_{\rm EA-stabilizer}$ as codes that can be encoded as follows:
\begin{align}
    \mathbb{Z}_2^{\otimes k}\xrightarrow{\text{GKP}}\mathbb{R}^{\otimes k}\xrightarrow{ \rm GKP\rhd EA-stabilizer}\mathbb{R}^{\otimes (n+c)}.
    \label{eq:ea_GKP_stab_encoding}
\end{align}
Here a single-mode GKP code is first produced, and subsequently encoded into a multi-mode GKP code using shared emodes. During the decoding, modes including part of the emodes are corrected for CV errors and decoded into a single-mode GKP code.

Codes here implement GKP-embedded EA-stabilizer encoding and decoding circuits. We discuss a simple example to illustrate how the encoding, error correction and decoding works here.
Here we define and demonstrate a $\mathcal{C}_{\rm GKP}\rhd\mathcal{C}_{[[3,1,3;2]]}$ code that requires $2$ emodes to suppress both position and momentum quadrature errors. First, both Alice and Bob pre-share two pairs of GKP Bell states, which are prepared by sending two copies of a two-mode canonical GKP state through a 50:50 beam splitter. Next, Alice encodes a single logical qubit into an oscillator mode using the GKP code. After that, the single logical oscillator mode is encoded into three oscillator modes using two emodes by applying the Gaussian encoder of the $[[3,1,3;2]]$ EA-repetition code. As $[[3,1,3;2]]$ is a maximal-entanglement EA-stabilizer code, the number of ancillary modes is zero. Thus, the logical GKP encoded state of the $\mathcal{C}_{\rm GKP}\rhd\mathcal{C}_{ [[3,1,3;2]]}$ code is \begin{align}
    \ket{\psi_{\rm L}}=(U_{\rm enc}^{G} \otimes I_{B})\left(\ket{\psi_{\rm GKP}} \otimes (\ket{\Phi^{+}_{\rm GKP}}_{AB}^{\otimes 2})\right).
    \label{eq:31EA-GKP-stabilizer}
\end{align}
Here we label the data mode by $1$ and $j\in\{2,3\}$, and $k\in\{4,5\}$ are Alice and Bob's parts of the emodes, respectively. The logical data mode is stabilized by two computational GKP stabilizers, given by the quadrature operators on the first mode conjugated by the $U_{\rm enc}^{G}$ i.e. by 
\begin{eqnarray}
    \hat{S}_q^{(1)} &= (U_{\rm enc}^{G}\otimes I_B)e^{i2\sqrt{\pi}\hat{q}_{1}}(U_{\rm enc}^{G}\otimes I_B )^{\dagger} \nonumber \\
    \hat{S}_p^{(1)} &= (U_{\rm enc}^{G}\otimes I_B )e^{-i2\sqrt{\pi}\hat{p}_{1}}(U_{\rm enc}^{G}\otimes I_B )^{\dagger}.
    \label{eq:stabilizer1defn}
\end{eqnarray}
Similarly, each of the two emodes in Alice's side is stabilized by two computational GKP stabilizers, given by the quadrature operators on the emodes conjugated by the $U_{\rm enc}^{G}\otimes I_B$ i.e. for $j\in\{2,3\}$, by
\begin{eqnarray}
\hat{S}_q^{(j)}&=& (U_{\rm enc}^{G}\otimes I_B )(e^{i2\sqrt{\pi}(\hat{q}_{j}+\hat{q}_{j+2})}) (U_{\rm enc}^{G}\otimes I_B )^{\dagger}\nonumber\\
\hat{S}_p^{(j)}&=& (U_{\rm enc}^{G}\otimes I_B )(e^{i2\sqrt{\pi}(\hat{p}_{j}+\hat{p}_{j+2})}) (U_{\rm enc}^{G}\otimes I_B )^{\dagger},
\label{eq:stabilizeremodesdefn}
\end{eqnarray}

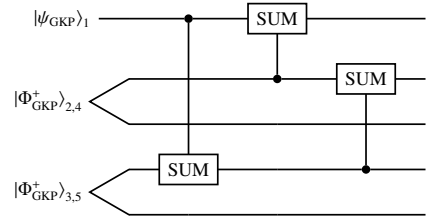
\begin{figure}
\centering
\begin{tikzpicture}
\node[scale=0.8] {
    \begin{quantikz}
\lstick{$\ket{\psi_{\rm GKP}}_{1}$}  &\qw &\ctrl{3} &\gate{\rm SUM} &\qw &\qw\\
& \makeebit[-60][]{$\ket{\Phi^{+}_{\rm GKP}}_{2,4}$} &  \qw &\ctrl{-1} &\gate{\rm SUM} &\qw\\
&     &\qw &\qw &\qw &\qw\\
& \makeebit[-60][]{$\ket{\Phi^{+}_{\rm GKP}}_{3,5}$} &\gate{\rm SUM} &\qw &\ctrl{-2} &\qw\\
&     &\qw &\qw &\qw &\qw
\end{quantikz}
};
\end{tikzpicture}
\caption{The Gaussian encoding circuit of $\mathcal{C}_{\rm GKP}\rhd\mathcal{C}_{[[3,1,3;2]]}$, which is formed when a computational GKP state is encoded into a multi-mode GKP code using the encoding circuit of $[[3,1,3;2]]_2$ code (Figure \ref{fig:encoding_circuit_3ea_rhd_GKP}).}
\label{fig:encoding_circuit_3ea-rep-GKP}
\end{figure}

The encoding circuit of the $\mathcal{C}_{\rm GKP}\rhd\mathcal{C}_{[[3,1,3;2]]}$ code is designed in Fig. \ref{fig:encoding_circuit_3ea-rep-GKP}. 
The corresponding Gaussian unitary encoder for the $[[3,1,3;2]]$ code is given by:
\begin{align}
    U_{\rm enc}^{G} &= \rm{SUM}_{3 \rightarrow 2}\; SUM_{2 \rightarrow 1}\; SUM_{1 \rightarrow 3} \nonumber \\
    &=\begin{pmatrix}
         1  &1  &0  &0  &0  &0\\
         1  &1  &1  &0  &0  &0\\
         1  &0  &1  &0  &0  &0\\
         0  &0  &0  &1  &0 &-1\\
         0  &0  &0 &-1  &1  &1\\
         0  &0  &0  &1 &-1  &0\\
    \end{pmatrix}.
    \label{eq:encoder_three_mode}
\end{align}
This may be noted as a GKP version of the encoded defined in Eq. (\ref{eq:uenc}). The action of the SUM gate on the quadratures is as follows \cite{xu2023qubit-oscillator}:
\begin{align}
    \text{SUM}_{i \rightarrow j} \hat{q}_i  \text{SUM}^{\dagger}_{i \rightarrow j}&=\hat{q}_i,\nonumber\\
    \text{SUM}_{i \rightarrow j}\hat{q}_j \text{SUM}^{\dagger}_{i \rightarrow j}&=\hat{q}_j-\hat{q}_i,\nonumber\\
    \text{SUM}_{i \rightarrow j}\hat{p}_i \text{SUM}^{\dagger}_{i \rightarrow j}&=\hat{p}_i+\hat{p}_j,\nonumber\\
    \text{SUM}_{i \rightarrow j}\hat{p}_j\text{SUM}^{\dagger}_{i \rightarrow j}&=\hat{p}_j.
\end{align}
The data mode is initialized with $\ket{\psi_{\rm GKP}}$, and the second and third modes in Alice's possession are initialized with her parts from the GKP Bell pairs. From Eq. (\ref{eq:encoder_three_mode}), it follows that under encoding and noiseless decoding, the data mode quadratures are transformed as:
\begin{align}
    q_1&\leftrightarrow q_1-q_2+q_3,\nonumber\\
    p_1&\leftrightarrow p_1+p_2+p_3.
    \label{eq:q_p_transformed}
\end{align}

From Eqs. (\ref{eq:stabilizer1defn}) and (\ref{eq:encoder_three_mode}), the $q$ and $p$ quadrature GKP stabilizers for the data mode are
\begin{align}
    \hat{S}_q^{(1)}&=e^{i2\sqrt{\pi}\left(\hat{q}_1-\hat{q}_2+\hat{q}_3 \right)}, \nonumber\\
    \hat{S}_p^{(1)}&=e^{-i2\sqrt{\pi}\left(\hat{p}_1+\hat{p}_2+\hat{p}_3 \right)},
    \label{eq:stab_three_data_mode}
\end{align}
and the corresponding measured syndromes for the data mode are
\begin{align}
     z_q^{(1)}&=R_{\sqrt{\pi}}(\zeta_{q}^{(1)}-\zeta_{q}^{(2)}+\zeta_{q}^{(3)}),\nonumber\\
     z_p^{(1)}&=R_{\sqrt{\pi}}(\zeta_{p}^{(1)}+\zeta_{p}^{(2)}+\zeta_{p}^{(3)}).
     \label{eq:syndromes_logical_1st_mode}
\end{align}
 From Eqs. (\ref{eq:stabilizeremodesdefn}) and (\ref{eq:encoder_three_mode}), the $q$ and $p$ quadrature GKP stabilizers for the emdoes in Alice's possession are
\begin{align}
    \hat{S}_q^{(2)}&=  e^{i\sqrt{\pi}(\hat{q}_2-\hat{q}_3 + \hat{q}_4)}, \nonumber\\
    \hat{S}_p^{(2)}&=e^{-i\sqrt{\pi}(\hat{p}_2+\hat{p}_1 + \hat{p}_4)},\nonumber\\
    \hat{S}_q^{(3)}&=e^{i\sqrt{\pi}(\hat{q}_2-\hat{q}_1+\hat{q}_5)}, \nonumber \\
    \hat{S}_p^{(3)}&=e^{-i\sqrt{\pi}(\hat{p}_3+\hat{p}_2 + \hat{p}_5)}.
    \label{eq:stab_three_emodes}
\end{align}
and the corresponding measured syndromes are
\begin{align}
    z_q^{(2)}&=R_{\sqrt{\pi}}(\zeta_{q}^{(2)}-\zeta_{q}^{(3)}),\nonumber\\
    z_p^{(2)}&=R_{\sqrt{\pi}}(\zeta_{p}^{(2)}+\zeta_{p}^{(1)}),\nonumber\\
    z_q^{(3)}&=R_{\sqrt{\pi}}(\zeta_{q}^{(2)}-\zeta_{q}^{(1)}),\nonumber\\
    z_p^{(3)}&=R_{\sqrt{\pi}}(\zeta_{p}^{(3)}+\zeta_{p}^{(2)}),
     \label{eq:syndromes_cano_3_mode}
\end{align}
where $\zeta_{q}^{(m)}$ and $\zeta_{p}^{(n)}$ $(m,n=4,5)$ both vanish by assumption of noiselessness of Bob's emodes.

From the above syndrome values, Eqs. (\ref{eq:syndromes_logical_1st_mode}) and (\ref{eq:syndromes_cano_3_mode}), the position and momentum quadrature errors are given by
 \begin{align}
    (\zeta_{q}^{(1)},\zeta_{q}^{(2)},\zeta_{q}^{(3)})&=(z_{q}^{(1)}+z_{q}^{(2)},z_q^{(1)}+z_{q}^{(2)}+z_{q}^{(3)}, z_q^{(1)}+z_q^{(3)}),\nonumber\\
(\zeta_{p}^{(1)},\zeta_{p}^{(2)},\zeta_{p}^{(3)})&=(z_p^{(1)}-z_p^{(3)},-z_p^{(1)}+z_p^{(2)}+z_p^{(3)}, z_{p}^{(1)}-z_{p}^{(2)}).
    \label{eq:zeta=}
 \end{align}
Here note that $z_q^{(1)}$ and $z_p^{(1)}$ are not measured, since such measurement would disturb the data mode. Thus each of the $q$ and $p$ shift errors in Eq. (\ref{eq:zeta=}) can only be known up to margin determined by the indeterminate $z_q^{(1)}$ and $z_p^{(1)}$. Therefore, to arrive at a deterministic value-assignment to $\zeta_q^{(1)}$ or $\zeta_p^{(1)}$, we require a further condition. Following Ref. \cite{noh2020encoding}, we use the maximum likelihood estimation formalism \cite{noh2020encoding, xu2023qubit-oscillator} to infer the most probable actual error from the syndrome measurement outcomes  $z_{q/p}^{(j)}$. Because all nosies are Gaussian and iid: $(\zeta_{q}^{(1)},\zeta_{q}^{(2)},\zeta_{q}^{(3)})\sim\mathcal{N}(0, \sigma^{2})$. The Gaussian noise strongly favors small shifts over large ones. Therefore, among all choices compatible with syndromes, the most probable choice is the one with the smallest total squared magnitude. In light of this property, the joint probability density is expressed as 
\begin{align}
    p_\sigma(\zeta_{q}^{(1)},\zeta_{q}^{(2)},\zeta_{q}^{(3)})\propto \exp \left(-\frac{({\zeta_{q}^{(1)}})^{2}+(\zeta_{q}^{(2)})^{2}+(\zeta_{q}^{(3)})^2}{2\sigma^2}\right).
    \label{eq:joint_prob}
\end{align}
Thus, to find most probable error value we maximize the probability Eq. (\ref{eq:joint_prob}), or equivalently minimize $ \sum_{j=1}^{3}({\zeta_{q}^{(j)}})^{2}$ and $ \sum_{j=1}^{3}({\zeta_{p}^{(j)}})^{2}$. Hence, the most probable position and momentum quadrature errors on the data mode are estimated as
\begin{align}
    \tilde{\zeta}_{q}^{(1)}&= {\rm{argmin}}_{z_{q}^{(1)}}\bigg[({\zeta_{q}^{(1)}})^{2}+(\zeta_{q}^{(2)})^{2}+(\zeta_{q}^{(3)})^{2}\bigg]\nonumber\\
    &=-\frac{2}{3}(z_{q}^{(2)}+z_{q}^{(3)}),\nonumber\\
    \tilde{\zeta}_{p}^{(1)}&= {\rm{argmin}}_{z_{p}^{(1)}}\bigg[({\zeta_{p}^{(1)}})^{2}+(\zeta_{p}^{(2)})^{2}+(\zeta_{p}^{(3)})^{2}\bigg]\nonumber\\
    &=\frac{2}{3}(z_{p}^{(2)}+z_{p}^{(3)}),
    \label{eq:estimated_error_rate_3_mode}
\end{align}
here $\text{argmin}_{x}f(x)$ are the values of $x$ for which $f(x)$ attains its minimum value. The pre-correction shift errors on the data mode, which can be determined by directly decoding the 3-mode state via Eq. (\ref{eq:q_p_transformed}), are found to be  ${z}_{q/p}^{(1)}$, the data mode syndromes themselves. Thus, from Eqs. (\ref{eq:syndromes_logical_1st_mode}) and (\ref{eq:estimated_error_rate_3_mode}), the residual (post-correction) logical position and momentum quadrature errors on the data mode are calculated as
\begin{align}
    \zeta_{q}^{(*)}=z_{q}^{(1)}-\tilde{\zeta}_{q}^{(1)}=\frac{1}{3}(\zeta_{q}^{(1)}+\zeta_{q}^{(2)}+\zeta_{q}^{(3)}),\nonumber\\
     \zeta_{p}^{(*)}=z_{p}^{(1)}-\tilde{\zeta}_{p}^{(1)}=\frac{1}{3}(\zeta_{p}^{(1)}-\zeta_{p}^{(2)}+\zeta_{p}^{(3)}),
     \label{eq:residual}
   \end{align}
here we assume that $z_{q/p}\in[-\sqrt{\pi}/2,\sqrt{\pi}/2]$ in order to remove $R_{\sqrt{\pi}}$. For independent variables $X_i$ with variances $V(X_i)$, the variance of their sum is the sum of their variances: $V(\sum_j X_j) = \sum_j V(X_j)$. Furthermore, the scaling of a variable produces a quadratic scaling of variance: $V(\alpha X) = \alpha^2 V(X)$. Moreover, it is known that the normal distribution is closed under linear transformations.

In light of these properties of probability distribution functions, since $(\zeta_{q}^{(1)},\zeta_{p}^{(1)},\zeta_{q}^{(2)},\zeta_{p}^{(2)},\zeta_{q}^{(3)},\zeta_{p}^{(3)})\sim_{\rm iid}\mathcal{N}(0, \sigma^{2})$, thus the variance of the residual logical position and momentum quadrature errors in the data mode are
\begin{align}
    \zeta_{q}^{(*)} \sim \mathcal{N}\bigg(0,\sigma_{q}^{2}=\frac{\sigma^{2}}{3}\bigg),\nonumber\\
    \zeta_{p}^{(*)} \sim \mathcal{N}\bigg(0,\sigma_{p}^{2}=\frac{\sigma^{2}}{3}\bigg),
    \label{eq:residual_supression-three_mode}
\end{align}
 which indicates that both quadrature error operators on the data mode are suppressed by a factor $(1/3)$. To show this explicitly, see Appendix \ref{sec:pdf}.

\begin{figure}[t]
    \centering
    \includegraphics[width=8cm]{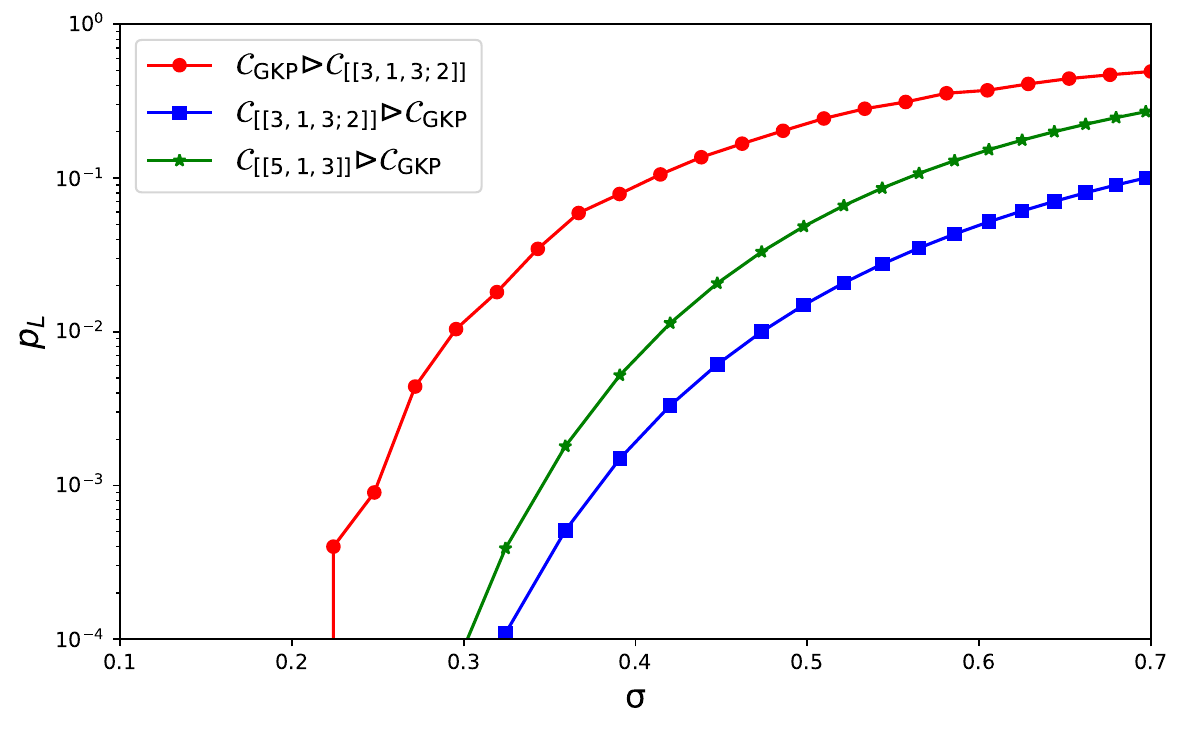}
    \caption{(Color online) Logical failure probability of $\mathcal{C}_{[[3,1,3;2]]}\rhd\mathcal{C}_{\rm GKP}$, $\mathcal{C}_{[[5,1,3]]}\rhd\mathcal{C}_{\rm GKP}$, and $\mathcal{C}_{\rm GKP}\rhd\mathcal{C}_{ [[3,1,3;2]]}$ codes as a function of the standard deviation $\sigma$, depicted as the blue, green, and blue plots, respectively.}
     \label{fig:logical_error_rate_3eagkp_gkp3ea_5qubit}
     \end{figure}

 Finally, we compare the performance of the $\mathcal{C}_{\rm GKP}\rhd\mathcal{C}_{[[3,1,3;2]]}$ code with the $\mathcal{C}_{[[3,1,3;2]]}\rhd\mathcal{C}_{\rm GKP}$ code by depicting their logical failure probability in Fig. \ref{fig:logical_error_rate_3eagkp_gkp3ea_5qubit}. We observe that the latter outperforms the former, which can be attributed to the fact that the former requires estimating unmeasured states, whereas the latter involves a full suite of EA stabilizer measurements. This result is similar to that fact, reported in Ref. \cite{xu2023qubit-oscillator}, that $\mathcal{C}_{\rm GKP}\rhd\mathcal{C}_{\rm stabilizer}$ codes fare worse than the corresponding $\mathcal{C}_{\rm stabilizer}\rhd\mathcal{C}_{\rm GKP}$ code, in the case of the stabilizer codes $[[5,1,3]]$, $[[7,1,3]]$, and $[[9,1,3]]$.

\section{GKP $\rhd$ $[[n,1,n;n-1]]$ EA-repetition codes}\label{sec:GKP_rhd_n_EA_rep}

Here we generalize the $\mathcal{C}_{\rm GKP}\rhd\mathcal{C}_{[[3,1,3;2]]}$ code to the $\mathcal{C}_{\rm GKP}\rhd\mathcal{C}_{[[n,1,n;n-1]]}$ code that is defined by specifically considering the encoder of the $[[n,1,n,n-1]]$ EA-repetition code for odd $n\ge 3$. The encoded state for the $\mathcal{C}_{\rm GKP}\rhd\mathcal{C}_{[[n,1,n;n-1]]}$ code is defined as:
\begin{align}
    \ket{\psi_{L}}=(U_{\rm enc}^{G}\otimes I_B)(\ket{\psi_{\rm GKP}}\otimes \ket{\rm \Phi_{\rm GKP}^{+}}_{AB}^{\otimes(n-1)}),
\end{align}
where $U_{\rm enc}^{G}$ is the relevant Gaussian encoder for the $\mathcal{C}_{\rm GKP}\rhd\mathcal{C}_{ [[n,1,n;n-1]]}$ code. Here, the first mode is data mode, and the rest $(n-1)$ emodes are initialized with GKP Bell pairs.

Analogous to Eqs. (\ref{eq:uenc}) and (\ref{eq:encoder_three_mode}), we can write encoders for the $[[n,1,n; n-1]]$ and corresponding $\mathcal{C}_{\rm GKP}\rhd\mathcal{C}_{[[n,1,n;n-1]]}$ codes. In the manner of the stabilizers Eq. (\ref{eq:stab_three_data_mode}) for the $\mathcal{C}_{\rm GKP}\rhd\mathcal{C}_{ [[3,1,3;2]]}$ code, the stabilizers for the data mode in the $\mathcal{C}_{\rm GKP}\rhd\mathcal{C}_{[[n,1,n;n-1]]}$ code are
\begin{align}
    \hat{S}_q^{(1)}&=e^{i2\sqrt{\pi}\big(\sum_{k=1}^{n}(-1)^{k+1}\hat{q}_{k}\big)},\\ \nonumber
    \hat{S}_p^{(1)}&=e^{-i2\sqrt{\pi}(\sum_{k=1}^{n}\hat{p}^{k})},
\end{align}
and the corresponding measured syndromes for the data mode are
\begin{align}
    z_q^{(1)}&=R_{\sqrt{\pi}}\bigg(\sum_{k=1}^{n}(-1)^{k+1}\zeta_{q}^{(k)}\bigg),\\ \nonumber
    z_p^{(1)}&=R_{\sqrt{\pi}}\bigg(\sum_{k=1}^{n}\zeta_{p}^{(k)}\bigg).
    \label{eq:syndromes_logical_n_mode}
\end{align}
where $k\in \mathbb{Z}^{+}$.
Similarly, in the manner of the stabilizers for the emodes Eq. (\ref{eq:stab_three_emodes}) for the $\mathcal{C}_{\rm GKP}\rhd\mathcal{C}_{[[3,1,3;2]]}$,  the GKP stabilizers for the parts of emodes in Alice's possession are in the $\mathcal{C}_{\rm GKP}\rhd\mathcal{C}_{ [[n,1,n;n-1]]}$ code are
\begin{align}
    \hat{S}_q^{(2k)}&=e^{i\sqrt{\pi}\big(\sum_{j=2k}^{2n-2}(-1)^{\frac{j}{2}+1}\hat{q}_{\frac{j}{2}+1}+\hat{q}_{2k+2}\big)}, \\ \nonumber
    \hat{S}_p^{(2k)}&=e^{-i\sqrt{\pi}\big(\hat{p}_{2k-1}+\hat{p}_{2k}+\hat{p}_{2k+2}\big)}, \\ \nonumber
    \hat{S}_q^{(2k+1)}&=e^{i\sqrt{\pi}\big(\hat{q}_{2k}-\hat{q}_{2k-1}+\hat{q}_{2k+3}\big)},\\ \nonumber
    \hat{S}_p^{(2k+1)}&=e^{-i\sqrt{\pi}\big(\sum_{j=2k}^{n}\hat{p}_{j}+\hat{p}_{2k+3}\big)}, 
\end{align}
where $j\in \{2,3,\cdots,n\}$, and corresponding measured syndromes for parts of emodes in Alice's possession are
\begin{align}
    z_q^{(2k)}&=R_{\sqrt{\pi}}\bigg(\sum_{j=2k}^{2n-2}(-1)^{\frac{j}{2}+1}\zeta_{q}^{(\frac{j}{2}+1)}\bigg),\\ \nonumber
    z_p^{(2k)}&=R_{\sqrt{\pi}}\bigg(\zeta_{\hat{p}}^{(2k-1)}+\zeta_{p}^{(2k)}\bigg),\\ \nonumber
    z_q^{(2k+1)}&=R_{\sqrt{\pi}}\bigg(\zeta_{\hat{q}}^{(2k)}-\zeta_{q}^{(2k-1)}\bigg), \\ \nonumber
    z_p^{(2k+1)}&=R_{\sqrt{\pi}}\bigg(\sum_{j=2k}^{n}\zeta_{p}^{(j)}\bigg),
    \label{eq:syndromes_cano_n_mode}
\end{align}
where $\zeta_{q}^{(2k+m)}$ and $\zeta_{p}^{(2k+m)}$ $(m=2,3)$ both vanish by assumption of noiselessness of Bob's emodes. 

Analogous to Eqs. (\ref{eq:zeta=}) and (\ref{eq:joint_prob}) we can write the quadrature errors and joint probability density for $\mathcal{C}_{\rm GKP}\rhd\mathcal{C}_{ [[n,1,n;n-1]]}$ code. In the manner of most probable quadrature errors Eq. (\ref{eq:estimated_error_rate_3_mode}) for the $\mathcal{C}_{\rm GKP}\rhd\mathcal{C}_{ [[3,1,3;2]]}$ code, the most probable position and momentum quadrature errors on the data mode for the $\mathcal{C}_{\rm GKP}\rhd\mathcal{C}_{[[n,1,n;n-1]]}$ code are calculated as
\begin{align}
    \tilde{\zeta}_{q}^{(1)}&= -\Big(\frac{n-1}{n}\Big)z_{q}^{(3)}-\Big(\frac{n-3}{n}\Big)z_{q}^{(5)}-\Big(\frac{n-5}{n}\Big)z_{q}^{(7)}- \cdots \nonumber\\
    &-\frac{2}{n}\big(z_{q}^{(2)}+z_{q}^{(4)}+z_{q}^{(6)}+\cdots +z_{q}^{(n-1)}+z_{q}^{(n)}\big),\nonumber\\
    \tilde{\zeta}_{p}^{(1)}&= \Big(\frac{n-1}{n}\Big)z_{p}^{(2)}+\Big(\frac{n-3}{n}\Big)z_{p}^{(4)}+\Big(\frac{n-5}{n}\Big)z_{p}^{(6)}+ \cdots \nonumber\\
    &+\frac{2}{n}\big(z_{p}^{(3)}+z_{p}^{(5)}+z_{p}^{(7)}+\cdots +z_{p}^{(n-1)}+z_{p}^{(n)}\big).
\end{align}
here we assume that $z_{q/p}\in[-\sqrt{\pi}/2,\sqrt{\pi}/2]$ in order to remove $R_{\sqrt{\pi}}$. For $n=3$, the above equations reduce to Eq. (\ref{eq:estimated_error_rate_3_mode}). The residual logical position and momentum quadrature errors of the data mode are
\begin{align}
    \zeta_{q}^{(*)}&=z_{q}^{(1)}-\tilde{\zeta}_{q}^{(1)}\nonumber\\
    &=\frac{1}{n}\bigg(\sum_{k=1}^{n}\zeta_{q}^{(n)}\bigg),\nonumber\\
     \zeta_{p}^{(*)}&=z_{p}^{(1)}-\tilde{\zeta}_{p}^{(1)}\nonumber\\
     &=\frac{1}{n}\bigg(\sum_{k=1}^{n}(-1)^{[k=2]}\zeta_{p}^{(n)}\bigg), \hspace{1cm}[k=2]{}= 
\begin{cases}
1 & \text{if } k = 2 \\
0 & \text{otherwise}
\end{cases}~~,
\label{eq:residual_n_modes}
   \end{align}
 Since, $(\zeta_{q}^{(1)},\zeta_{p}^{(1)},\zeta_{q}^{(2)},\zeta_{p}^{(2)},\cdots,\zeta_{q}^{(n)},\zeta_{p}^{(n)})\sim_{\rm iid}\mathcal{N}(0, \sigma^{2})$, thus the variance of the residual logical position and momentum quadrature errors in the data mode are
\begin{align}
    \zeta_{q}^{(*)} \sim \mathcal{N}\bigg(0,\sigma_{q}^{2}=\frac{1}{n}\sigma^2\bigg),\nonumber\\
    \zeta_{p}^{(*)} \sim \mathcal{N}\bigg(0,\sigma_{p}^{2}=\frac{1}{n}\sigma^2\bigg).
\end{align}
Thus, the variance of both position and momentum quadrature errors is suppressed by a factor $(1/n)$. Thus, we have the following theorem.
\begin{theorem}
An $\mathcal{C}_{\rm GKP}\rhd\mathcal{C}_{\rm EA-repetition}$ concatenated code constructed by implementing GKP-embedded EA-stabilizer of the $[[n,1,n;n-1]]$ EA-repetition code, suppresses variance of position and momentum quadrature errors by a factor $(1/n)$ with the use of $n-1$ emodes.
\label{thm:nmodesuppression}
\end{theorem}

An unbiased $(2n+1)$-mode GKP-repetition code proposed in Ref. \cite{xu2023qubit-oscillator} that scales down both quadrature errors by a factor $1/(n+1)$ with the requirement of $2n$ ancillary modes. Note that by Theorem \ref{thm:nmodesuppression}, a $\mathcal{C}_{\rm GKP}\rhd\mathcal{C}_{[[n+1,1,n+1;n]]}$, which uses $2n+1$ total number of modes, produces a similar suppression by a factor of $1/(n+1)$.

\section{Conclusions}\label{sec:conclusions}

In this work, we present two approaches for concatenating entanglement-assisted (EA) stabilizer codes with Gottesman–Kitaev–Preskill (GKP) codes. The first approach uses an EA-stabilizer outer code concatenated with a GKP inner code; the second uses a GKP outer code concatenated with an EA-stabilizer inner code. As an example of the first approach, we present a three-qubit EA-repetition code concatenated with a GKP code, which outperforms the five-qubit perfect code concatenated with a GKP code. As an example of the second approach, we present a GKP code concatenated with a three-qubit EA-repetition code, which suppresses errors in both quadratures of the data mode using two ebits. We generalize the latter example to a family of GKP codes concatenated with an $n$-qubit EA-repetition code, which suppresses errors in both quadratures by reducing their variances by a factor of $1/n$.

We have presented the encoding process, stabilizers, measured syndromes, and encoding circuits for these EA continuous-variable (CV) concatenated codes, and studied their performance by evaluating logical failure probabilities. Our analysis considers only Gaussian shift errors; however, other physical channels, such as the bosonic pure-loss channel \cite{cerf2007quantum}, can be accounted for, and the performance of our EA-CV concatenated codes can be tested accordingly. We have also assumed ideal GKP states; applying the Fock damping operator to them yields finite-energy GKP states, and our results can be extended to this finite-energy case.

The concatenation of qubit stabilizer and GKP codes is well studied for applications such as quantum repeaters \cite{rozpkedek2021quantum}. It would be interesting to incorporate both of our EA-CV concatenated codes into such communication settings to analyze their performance. Furthermore, catalytic stabilizer codes are relevant to the time evolution of quantum computation \cite{brun2014catalytic}. Therefore, finding a catalytic version of EA-CV concatenated codes for computational settings is a promising future direction. Beyond these, several other avenues merit exploration. First, experimental realization of our schemes can be pursued using trapped ions, superconducting circuits, or photonic platforms capable of generating both GKP states and entanglement-assisted stabilizer codes. Second, the resource overhead — particularly the number of ebits and oscillator modes — could be systematically optimized against the achieved logical failure probability reduction. Third, our concatenation frameworks may be extended to biased-noise channels \cite{li2024correcting, puri2019stabilized}, where the asymmetry between position and momentum quadrature errors can be exploited for further performance gains. Fourth, while we focus on GKP codes as the inner or outer code, other bosonic codes (e.g., cat codes \cite{cochrane1999cat, hastrup2022cat}, binomial codes \cite{michael2016binomial, laha2026binomial}) could replace the GKP layer within the same entanglement-assisted concatenation architecture, potentially offering better hardware-specific trade-offs. Finally, analyzing the fault-tolerant threshold of these EA-CV concatenated codes under realistic noise models-- including finite squeezing, loss, and imperfect entangling gates-- remains an open but critical direction toward practical quantum computing and communication.

\acknowledgments
N.R.D. thanks Jai Lalita and V.B. Sabale for helpful discussions on GKP states. N.R.D. acknowledges financial support from the Department of Science and Technology, Ministry of Science and Technology, India, through the INSPIRE fellowship. N.R.D also acknowledges support from the Indian Space Research Organisation (ISRO) project No. ISRO/RES/3/906/22-23.

\appendix
\section{Probability density function via convolution}\label{sec:pdf}
For completeness, we note that the probability density functions (PDFs) of the residual uncorrected errors can be determined directly, instead of the statistical identities we employed above. By using Eq. (\ref{eq:residual}), the PDFs of the logical quadrature errors $\zeta_q^{(*)}$ is calculated as the convolution:
 \begin{eqnarray}
     P(\zeta_q^{(*)}) &=& 3\iint_{-\infty}^{\infty}d\zeta_q^{(1)} d\zeta_q^{(2)} P\sigma(\zeta_q^{(1)})P(\zeta_q^{(2)})P(3\zeta^{(*)} - \zeta_q^{(1)} - \zeta_q^{(2)}), \nonumber \\
     &=& 3 \mathfrak{N}^3 \iint_{-\infty}^{\infty} e^{ -\frac{1}{2\sigma^2}\left(\zeta_q^{(1)2} + \zeta_q^{(2)2} + \left(3\zeta_q^{(*)}-\zeta_q^{(1)}-\zeta_q^{(2)}\right)^2\right)} \, d\zeta_q^{(1)} \, d\zeta_q^{(2)},
 \end{eqnarray}
where $\mathfrak{N}=\frac{1}{\sigma\sqrt{2\pi}}$ is the normalization factor coming from each integral.

To evaluate the double integral, one expands the exponent term: $\zeta_q^{(1)2} + \zeta_q^{(2)2} + \left(3g-\zeta_q^{(1)}-\zeta_q^{(2)}\right)^2$. We complete the square first with respect to $\zeta_q^{(1)}$, and evaluate that Gaussian integral over $d\zeta_q^{(1)}$. This will eliminate $\zeta_q^{(1)}$ and leave us with an expression containing only variable $\zeta_q^{(2)}$ and $\zeta_q^{(*)}$. We complete the square again with respect to $\zeta_q^{(2)}$, and we then evaluate the remaining Gaussian integral over $d\zeta_q^{(2)}$. Once the constants are simplified outside the integral, we will find that we arrive at the exact same result: a Gaussian with a mean of 0 and a variance of $\frac{\sigma^2}{3}$. The linear combination algebraic identities about iid's and Gaussian functions used in the main text saves us this extensive integral and algebraic bookkeeping. A similar exercise can be done for determining the PDF of $\zeta_p^{(*)}$. \color{black}

\bibliography{reference}

\begin{thebibliography}{52}%
\makeatletter
\providecommand \@ifxundefined [1]{%
 \@ifx{#1\undefined}
}%
\providecommand \@ifnum [1]{%
 \ifnum #1\expandafter \@firstoftwo
 \else \expandafter \@secondoftwo
 \fi
}%
\providecommand \@ifx [1]{%
 \ifx #1\expandafter \@firstoftwo
 \else \expandafter \@secondoftwo
 \fi
}%
\providecommand \natexlab [1]{#1}%
\providecommand \enquote  [1]{``#1''}%
\providecommand \bibnamefont  [1]{#1}%
\providecommand \bibfnamefont [1]{#1}%
\providecommand \citenamefont [1]{#1}%
\providecommand \href@noop [0]{\@secondoftwo}%
\providecommand \href [0]{\begingroup \@sanitize@url \@href}%
\providecommand \@href[1]{\@@startlink{#1}\@@href}%
\providecommand \@@href[1]{\endgroup#1\@@endlink}%
\providecommand \@sanitize@url [0]{\catcode `\\12\catcode `\$12\catcode `\&12\catcode `\#12\catcode `\^12\catcode `\_12\catcode `\%12\relax}%
\providecommand \@@startlink[1]{}%
\providecommand \@@endlink[0]{}%
\providecommand \url  [0]{\begingroup\@sanitize@url \@url }%
\providecommand \@url [1]{\endgroup\@href {#1}{\urlprefix }}%
\providecommand \urlprefix  [0]{URL }%
\providecommand \Eprint [0]{\href }%
\providecommand \doibase [0]{https://doi.org/}%
\providecommand \selectlanguage [0]{\@gobble}%
\providecommand \bibinfo  [0]{\@secondoftwo}%
\providecommand \bibfield  [0]{\@secondoftwo}%
\providecommand \translation [1]{[#1]}%
\providecommand \BibitemOpen [0]{}%
\providecommand \bibitemStop [0]{}%
\providecommand \bibitemNoStop [0]{.\EOS\space}%
\providecommand \EOS [0]{\spacefactor3000\relax}%
\providecommand \BibitemShut  [1]{\csname bibitem#1\endcsname}%
\let\auto@bib@innerbib\@empty
%</preamble>
\bibitem [{\citenamefont {Gottesman}(1997)}]{gottesman1997stabilizer}%
  \BibitemOpen
  \bibfield  {author} {\bibinfo {author} {\bibfnamefont {D.}~\bibnamefont {Gottesman}},\ }\emph {\bibinfo {title} {Stabilizer codes and quantum error correction}},\ \href@noop {} {Ph.D. thesis},\ \bibinfo  {school} {California Institute of Technology} (\bibinfo {year} {1997})\BibitemShut {NoStop}%
\bibitem [{\citenamefont {Lidar}\ and\ \citenamefont {Brun}(2013)}]{lidar2013quantum}%
  \BibitemOpen
  \bibfield  {author} {\bibinfo {author} {\bibfnamefont {D.~A.}\ \bibnamefont {Lidar}}\ and\ \bibinfo {author} {\bibfnamefont {T.~A.}\ \bibnamefont {Brun}},\ }\href {https://doi.org/10.1017/CBO9781139034807} {\emph {\bibinfo {title} {Quantum Error Correction}}}\ (\bibinfo  {publisher} {Cambridge University Press},\ \bibinfo {year} {2013})\BibitemShut {NoStop}%
\bibitem [{\citenamefont {Laflamme}\ \emph {et~al.}(1996)\citenamefont {Laflamme}, \citenamefont {Miquel}, \citenamefont {Paz},\ and\ \citenamefont {Zurek}}]{Laflamme1996perfect}%
  \BibitemOpen
  \bibfield  {author} {\bibinfo {author} {\bibfnamefont {R.}~\bibnamefont {Laflamme}}, \bibinfo {author} {\bibfnamefont {C.}~\bibnamefont {Miquel}}, \bibinfo {author} {\bibfnamefont {J.~P.}\ \bibnamefont {Paz}},\ and\ \bibinfo {author} {\bibfnamefont {W.~H.}\ \bibnamefont {Zurek}},\ }\bibfield  {title} {\bibinfo {title} {Perfect quantum error correcting code},\ }\href {https://doi.org/10.1103/PhysRevLett.77.198} {\bibfield  {journal} {\bibinfo  {journal} {Phys. Rev. Lett.}\ }\textbf {\bibinfo {volume} {77}},\ \bibinfo {pages} {198} (\bibinfo {year} {1996})}\BibitemShut {NoStop}%
\bibitem [{\citenamefont {Shor}(1995)}]{shor1995scheme}%
  \BibitemOpen
  \bibfield  {author} {\bibinfo {author} {\bibfnamefont {P.~W.}\ \bibnamefont {Shor}},\ }\bibfield  {title} {\bibinfo {title} {Scheme for reducing decoherence in quantum computer memory},\ }\href {https://doi.org/10.1103/PhysRevA.52.R2493} {\bibfield  {journal} {\bibinfo  {journal} {Phys. Rev. A}\ }\textbf {\bibinfo {volume} {52}},\ \bibinfo {pages} {R2493} (\bibinfo {year} {1995})}\BibitemShut {NoStop}%
\bibitem [{\citenamefont {Steane}(1996)}]{steane1996error}%
  \BibitemOpen
  \bibfield  {author} {\bibinfo {author} {\bibfnamefont {A.~M.}\ \bibnamefont {Steane}},\ }\bibfield  {title} {\bibinfo {title} {Error correcting codes in quantum theory},\ }\href {https://doi.org/10.1103/PhysRevLett.77.793} {\bibfield  {journal} {\bibinfo  {journal} {Phys. Rev. Lett.}\ }\textbf {\bibinfo {volume} {77}},\ \bibinfo {pages} {793} (\bibinfo {year} {1996})}\BibitemShut {NoStop}%
\bibitem [{\citenamefont {Braunstein}(1998)}]{Braunstein1998error}%
  \BibitemOpen
  \bibfield  {author} {\bibinfo {author} {\bibfnamefont {S.~L.}\ \bibnamefont {Braunstein}},\ }\bibfield  {title} {\bibinfo {title} {Error correction for continuous quantum variables},\ }\href {https://doi.org/10.1103/PhysRevLett.80.4084} {\bibfield  {journal} {\bibinfo  {journal} {Phys. Rev. Lett.}\ }\textbf {\bibinfo {volume} {80}},\ \bibinfo {pages} {4084} (\bibinfo {year} {1998})}\BibitemShut {NoStop}%
\bibitem [{\citenamefont {Lloyd}\ and\ \citenamefont {Slotine}(1998)}]{lloyd1998analog}%
  \BibitemOpen
  \bibfield  {author} {\bibinfo {author} {\bibfnamefont {S.}~\bibnamefont {Lloyd}}\ and\ \bibinfo {author} {\bibfnamefont {J.-J.~E.}\ \bibnamefont {Slotine}},\ }\bibfield  {title} {\bibinfo {title} {Analog quantum error correction},\ }\href {https://doi.org/10.1103/PhysRevLett.80.4088} {\bibfield  {journal} {\bibinfo  {journal} {Phys. Rev. Lett.}\ }\textbf {\bibinfo {volume} {80}},\ \bibinfo {pages} {4088} (\bibinfo {year} {1998})}\BibitemShut {NoStop}%
\bibitem [{\citenamefont {Lee}\ and\ \citenamefont {Jeong}(2013)}]{lee2013near}%
  \BibitemOpen
  \bibfield  {author} {\bibinfo {author} {\bibfnamefont {S.-W.}\ \bibnamefont {Lee}}\ and\ \bibinfo {author} {\bibfnamefont {H.}~\bibnamefont {Jeong}},\ }\bibfield  {title} {\bibinfo {title} {Near-deterministic quantum teleportation and resource-efficient quantum computation using linear optics and hybrid qubits},\ }\href {https://doi.org/10.1103/PhysRevA.87.022326} {\bibfield  {journal} {\bibinfo  {journal} {Phys. Rev. A}\ }\textbf {\bibinfo {volume} {87}},\ \bibinfo {pages} {022326} (\bibinfo {year} {2013})}\BibitemShut {NoStop}%
\bibitem [{\citenamefont {Kapit}(2016)}]{kapit2016hardware}%
  \BibitemOpen
  \bibfield  {author} {\bibinfo {author} {\bibfnamefont {E.}~\bibnamefont {Kapit}},\ }\bibfield  {title} {\bibinfo {title} {Hardware-efficient and fully autonomous quantum error correction in superconducting circuits},\ }\href {https://doi.org/10.1103/PhysRevLett.116.150501} {\bibfield  {journal} {\bibinfo  {journal} {Phys. Rev. Lett.}\ }\textbf {\bibinfo {volume} {116}},\ \bibinfo {pages} {150501} (\bibinfo {year} {2016})}\BibitemShut {NoStop}%
\bibitem [{\citenamefont {Liu}\ \emph {et~al.}(2026)\citenamefont {Liu}, \citenamefont {Singh}, \citenamefont {Smith}, \citenamefont {Crane}, \citenamefont {Martyn}, \citenamefont {Eickbusch}, \citenamefont {Schuckert}, \citenamefont {Li}, \citenamefont {Sinanan-Singh}, \citenamefont {Soley}, \citenamefont {Tsunoda}, \citenamefont {Chuang}, \citenamefont {Wiebe},\ and\ \citenamefont {Girvin}}]{liu2026hybrid}%
  \BibitemOpen
  \bibfield  {author} {\bibinfo {author} {\bibfnamefont {Y.}~\bibnamefont {Liu}}, \bibinfo {author} {\bibfnamefont {S.}~\bibnamefont {Singh}}, \bibinfo {author} {\bibfnamefont {K.~C.}\ \bibnamefont {Smith}}, \bibinfo {author} {\bibfnamefont {E.}~\bibnamefont {Crane}}, \bibinfo {author} {\bibfnamefont {J.~M.}\ \bibnamefont {Martyn}}, \bibinfo {author} {\bibfnamefont {A.}~\bibnamefont {Eickbusch}}, \bibinfo {author} {\bibfnamefont {A.}~\bibnamefont {Schuckert}}, \bibinfo {author} {\bibfnamefont {R.~D.}\ \bibnamefont {Li}}, \bibinfo {author} {\bibfnamefont {J.}~\bibnamefont {Sinanan-Singh}}, \bibinfo {author} {\bibfnamefont {M.~B.}\ \bibnamefont {Soley}}, \bibinfo {author} {\bibfnamefont {T.}~\bibnamefont {Tsunoda}}, \bibinfo {author} {\bibfnamefont {I.~L.}\ \bibnamefont {Chuang}}, \bibinfo {author} {\bibfnamefont {N.}~\bibnamefont {Wiebe}},\ and\ \bibinfo {author} {\bibfnamefont {S.~M.}\ \bibnamefont {Girvin}},\ }\bibfield  {title} {\bibinfo {title} {Hybrid oscillator-qubit quantum processors: Instruction set
  architectures, abstract machine models, and applications},\ }\href {https://doi.org/10.1103/4rf7-9tfx} {\bibfield  {journal} {\bibinfo  {journal} {PRX Quantum}\ }\textbf {\bibinfo {volume} {7}},\ \bibinfo {pages} {010201} (\bibinfo {year} {2026})}\BibitemShut {NoStop}%
\bibitem [{\citenamefont {Gottesman}\ \emph {et~al.}(2001)\citenamefont {Gottesman}, \citenamefont {Kitaev},\ and\ \citenamefont {Preskill}}]{gottesman2001encoding}%
  \BibitemOpen
  \bibfield  {author} {\bibinfo {author} {\bibfnamefont {D.}~\bibnamefont {Gottesman}}, \bibinfo {author} {\bibfnamefont {A.}~\bibnamefont {Kitaev}},\ and\ \bibinfo {author} {\bibfnamefont {J.}~\bibnamefont {Preskill}},\ }\bibfield  {title} {\bibinfo {title} {Encoding a qubit in an oscillator},\ }\href {https://doi.org/10.1103/PhysRevA.64.012310} {\bibfield  {journal} {\bibinfo  {journal} {Phys. Rev. A}\ }\textbf {\bibinfo {volume} {64}},\ \bibinfo {pages} {012310} (\bibinfo {year} {2001})}\BibitemShut {NoStop}%
\bibitem [{\citenamefont {Grimsmo}\ and\ \citenamefont {Puri}(2021)}]{grimsmo2021quantum}%
  \BibitemOpen
  \bibfield  {author} {\bibinfo {author} {\bibfnamefont {A.~L.}\ \bibnamefont {Grimsmo}}\ and\ \bibinfo {author} {\bibfnamefont {S.}~\bibnamefont {Puri}},\ }\bibfield  {title} {\bibinfo {title} {Quantum error correction with the gottesman-kitaev-preskill code},\ }\href {https://doi.org/10.1103/PRXQuantum.2.020101} {\bibfield  {journal} {\bibinfo  {journal} {PRX Quantum}\ }\textbf {\bibinfo {volume} {2}},\ \bibinfo {pages} {020101} (\bibinfo {year} {2021})}\BibitemShut {NoStop}%
\bibitem [{\citenamefont {Fukui}\ \emph {et~al.}(2017)\citenamefont {Fukui}, \citenamefont {Tomita},\ and\ \citenamefont {Okamoto}}]{fukui2017analog}%
  \BibitemOpen
  \bibfield  {author} {\bibinfo {author} {\bibfnamefont {K.}~\bibnamefont {Fukui}}, \bibinfo {author} {\bibfnamefont {A.}~\bibnamefont {Tomita}},\ and\ \bibinfo {author} {\bibfnamefont {A.}~\bibnamefont {Okamoto}},\ }\bibfield  {title} {\bibinfo {title} {Analog quantum error correction with encoding a qubit into an oscillator},\ }\href {https://doi.org/10.1103/PhysRevLett.119.180507} {\bibfield  {journal} {\bibinfo  {journal} {Phys. Rev. Lett.}\ }\textbf {\bibinfo {volume} {119}},\ \bibinfo {pages} {180507} (\bibinfo {year} {2017})}\BibitemShut {NoStop}%
\bibitem [{\citenamefont {Wang}(2019)}]{wang2019quantumerrorcorrectionGKP}%
  \BibitemOpen
  \bibfield  {author} {\bibinfo {author} {\bibfnamefont {Y.}~\bibnamefont {Wang}},\ }\href {https://arxiv.org/abs/1908.00147} {\bibinfo {title} {Quantum error correction with the gkp code and concatenation with stabilizer codes}} (\bibinfo {year} {2019}),\ \Eprint {https://arxiv.org/abs/1908.00147} {arXiv:1908.00147 [quant-ph]} \BibitemShut {NoStop}%
\bibitem [{\citenamefont {Lin}\ and\ \citenamefont {Noh}(2025)}]{lin2025exploring}%
  \BibitemOpen
  \bibfield  {author} {\bibinfo {author} {\bibfnamefont {M.}~\bibnamefont {Lin}}\ and\ \bibinfo {author} {\bibfnamefont {K.}~\bibnamefont {Noh}},\ }\bibfield  {title} {\bibinfo {title} {Exploring the quantum capacity of a gaussian random-displacement channel using gottesman-kitaev-preskill codes and maximum-likelihood decoding},\ }\href {https://doi.org/10.1103/PhysRevA.111.052445} {\bibfield  {journal} {\bibinfo  {journal} {Phys. Rev. A}\ }\textbf {\bibinfo {volume} {111}},\ \bibinfo {pages} {052445} (\bibinfo {year} {2025})}\BibitemShut {NoStop}%
\bibitem [{\citenamefont {Fukui}\ \emph {et~al.}(2018{\natexlab{a}})\citenamefont {Fukui}, \citenamefont {Tomita},\ and\ \citenamefont {Okamoto}}]{fukui2018tracking}%
  \BibitemOpen
  \bibfield  {author} {\bibinfo {author} {\bibfnamefont {K.}~\bibnamefont {Fukui}}, \bibinfo {author} {\bibfnamefont {A.}~\bibnamefont {Tomita}},\ and\ \bibinfo {author} {\bibfnamefont {A.}~\bibnamefont {Okamoto}},\ }\bibfield  {title} {\bibinfo {title} {Tracking quantum error correction},\ }\href {https://doi.org/10.1103/PhysRevA.98.022326} {\bibfield  {journal} {\bibinfo  {journal} {Phys. Rev. A}\ }\textbf {\bibinfo {volume} {98}},\ \bibinfo {pages} {022326} (\bibinfo {year} {2018}{\natexlab{a}})}\BibitemShut {NoStop}%
\bibitem [{\citenamefont {Fukui}\ \emph {et~al.}(2018{\natexlab{b}})\citenamefont {Fukui}, \citenamefont {Tomita}, \citenamefont {Okamoto},\ and\ \citenamefont {Fujii}}]{fukui2018highthreshold}%
  \BibitemOpen
  \bibfield  {author} {\bibinfo {author} {\bibfnamefont {K.}~\bibnamefont {Fukui}}, \bibinfo {author} {\bibfnamefont {A.}~\bibnamefont {Tomita}}, \bibinfo {author} {\bibfnamefont {A.}~\bibnamefont {Okamoto}},\ and\ \bibinfo {author} {\bibfnamefont {K.}~\bibnamefont {Fujii}},\ }\bibfield  {title} {\bibinfo {title} {High-threshold fault-tolerant quantum computation with analog quantum error correction},\ }\href {https://doi.org/10.1103/PhysRevX.8.021054} {\bibfield  {journal} {\bibinfo  {journal} {Phys. Rev. X}\ }\textbf {\bibinfo {volume} {8}},\ \bibinfo {pages} {021054} (\bibinfo {year} {2018}{\natexlab{b}})}\BibitemShut {NoStop}%
\bibitem [{\citenamefont {Vuillot}\ \emph {et~al.}(2019)\citenamefont {Vuillot}, \citenamefont {Asasi}, \citenamefont {Wang}, \citenamefont {Pryadko},\ and\ \citenamefont {Terhal}}]{vuillot2019quantum}%
  \BibitemOpen
  \bibfield  {author} {\bibinfo {author} {\bibfnamefont {C.}~\bibnamefont {Vuillot}}, \bibinfo {author} {\bibfnamefont {H.}~\bibnamefont {Asasi}}, \bibinfo {author} {\bibfnamefont {Y.}~\bibnamefont {Wang}}, \bibinfo {author} {\bibfnamefont {L.~P.}\ \bibnamefont {Pryadko}},\ and\ \bibinfo {author} {\bibfnamefont {B.~M.}\ \bibnamefont {Terhal}},\ }\bibfield  {title} {\bibinfo {title} {Quantum error correction with the toric gottesman-kitaev-preskill code},\ }\href {https://doi.org/10.1103/PhysRevA.99.032344} {\bibfield  {journal} {\bibinfo  {journal} {Phys. Rev. A}\ }\textbf {\bibinfo {volume} {99}},\ \bibinfo {pages} {032344} (\bibinfo {year} {2019})}\BibitemShut {NoStop}%
\bibitem [{\citenamefont {Noh}\ and\ \citenamefont {Chamberland}(2020)}]{noh2020faulttolerant}%
  \BibitemOpen
  \bibfield  {author} {\bibinfo {author} {\bibfnamefont {K.}~\bibnamefont {Noh}}\ and\ \bibinfo {author} {\bibfnamefont {C.}~\bibnamefont {Chamberland}},\ }\bibfield  {title} {\bibinfo {title} {Fault-tolerant bosonic quantum error correction with the surface--gottesman-kitaev-preskill code},\ }\href {https://doi.org/10.1103/PhysRevA.101.012316} {\bibfield  {journal} {\bibinfo  {journal} {Phys. Rev. A}\ }\textbf {\bibinfo {volume} {101}},\ \bibinfo {pages} {012316} (\bibinfo {year} {2020})}\BibitemShut {NoStop}%
\bibitem [{\citenamefont {Noh}\ \emph {et~al.}(2022)\citenamefont {Noh}, \citenamefont {Chamberland},\ and\ \citenamefont {Brand\~ao}}]{noh2022lowoverhead}%
  \BibitemOpen
  \bibfield  {author} {\bibinfo {author} {\bibfnamefont {K.}~\bibnamefont {Noh}}, \bibinfo {author} {\bibfnamefont {C.}~\bibnamefont {Chamberland}},\ and\ \bibinfo {author} {\bibfnamefont {F.~G.}\ \bibnamefont {Brand\~ao}},\ }\bibfield  {title} {\bibinfo {title} {Low-overhead fault-tolerant quantum error correction with the surface-gkp code},\ }\href {https://doi.org/10.1103/PRXQuantum.3.010315} {\bibfield  {journal} {\bibinfo  {journal} {PRX Quantum}\ }\textbf {\bibinfo {volume} {3}},\ \bibinfo {pages} {010315} (\bibinfo {year} {2022})}\BibitemShut {NoStop}%
\bibitem [{\citenamefont {Raveendran}\ \emph {et~al.}(2022)\citenamefont {Raveendran}, \citenamefont {Rengaswamy}, \citenamefont {Rozpedek}, \citenamefont {Raina}, \citenamefont {Jiang},\ and\ \citenamefont {Vasic}}]{Raveendran2022finiterateqldpcGKP}%
  \BibitemOpen
  \bibfield  {author} {\bibinfo {author} {\bibfnamefont {N.}~\bibnamefont {Raveendran}}, \bibinfo {author} {\bibfnamefont {N.}~\bibnamefont {Rengaswamy}}, \bibinfo {author} {\bibfnamefont {F.}~\bibnamefont {Rozpedek}}, \bibinfo {author} {\bibfnamefont {A.}~\bibnamefont {Raina}}, \bibinfo {author} {\bibfnamefont {L.}~\bibnamefont {Jiang}},\ and\ \bibinfo {author} {\bibfnamefont {B.}~\bibnamefont {Vasic}},\ }\bibfield  {title} {\bibinfo {title} {Finite {R}ate {QLDPC}-{GKP} {C}oding {S}cheme that {S}urpasses the {CSS} {H}amming {B}ound},\ }\href {https://doi.org/10.22331/q-2022-07-20-767} {\bibfield  {journal} {\bibinfo  {journal} {{Quantum}}\ }\textbf {\bibinfo {volume} {6}},\ \bibinfo {pages} {767} (\bibinfo {year} {2022})}\BibitemShut {NoStop}%
\bibitem [{\citenamefont {Xiao}\ and\ \citenamefont {Chen}(2022)}]{xiao2022quantum}%
  \BibitemOpen
  \bibfield  {author} {\bibinfo {author} {\bibfnamefont {H.}~\bibnamefont {Xiao}}\ and\ \bibinfo {author} {\bibfnamefont {X.}~\bibnamefont {Chen}},\ }\bibfield  {title} {\bibinfo {title} {Quantum convolutional codes concatenated with the {GKP} code for correcting continuous errors},\ }\href {https://doi.org/10.1007/s11128-022-03545-2} {\bibfield  {journal} {\bibinfo  {journal} {Quantum Inf. Process.}\ }\textbf {\bibinfo {volume} {21}},\ \bibinfo {pages} {Paper No. 198, 15} (\bibinfo {year} {2022})}\BibitemShut {NoStop}%
\bibitem [{\citenamefont {Li}\ and\ \citenamefont {Su}(2024)}]{li2024correcting}%
  \BibitemOpen
  \bibfield  {author} {\bibinfo {author} {\bibfnamefont {Z.}~\bibnamefont {Li}}\ and\ \bibinfo {author} {\bibfnamefont {D.}~\bibnamefont {Su}},\ }\bibfield  {title} {\bibinfo {title} {Correcting biased noise using gottesman-kitaev-preskill repetition code with noisy ancilla},\ }\href {https://doi.org/10.1103/PhysRevA.109.052420} {\bibfield  {journal} {\bibinfo  {journal} {Phys. Rev. A}\ }\textbf {\bibinfo {volume} {109}},\ \bibinfo {pages} {052420} (\bibinfo {year} {2024})}\BibitemShut {NoStop}%
\bibitem [{\citenamefont {Rozpedek}\ \emph {et~al.}(2021)\citenamefont {Rozpedek}, \citenamefont {Noh}, \citenamefont {Xu}, \citenamefont {Guha},\ and\ \citenamefont {Jiang}}]{rozpkedek2021quantum}%
  \BibitemOpen
  \bibfield  {author} {\bibinfo {author} {\bibfnamefont {F.}~\bibnamefont {Rozpedek}}, \bibinfo {author} {\bibfnamefont {K.}~\bibnamefont {Noh}}, \bibinfo {author} {\bibfnamefont {Q.}~\bibnamefont {Xu}}, \bibinfo {author} {\bibfnamefont {S.}~\bibnamefont {Guha}},\ and\ \bibinfo {author} {\bibfnamefont {L.}~\bibnamefont {Jiang}},\ }\bibfield  {title} {\bibinfo {title} {Quantum repeaters based on concatenated bosonic and discrete-variable quantum codes},\ }\href {https://doi.org/10.1038/s41534-021-00438-7} {\bibfield  {journal} {\bibinfo  {journal} {npj Quantum Information}\ }\textbf {\bibinfo {volume} {7}},\ \bibinfo {pages} {102} (\bibinfo {year} {2021})}\BibitemShut {NoStop}%
\bibitem [{\citenamefont {Chamberland}\ \emph {et~al.}(2022)\citenamefont {Chamberland}, \citenamefont {Noh}, \citenamefont {Arrangoiz-Arriola}, \citenamefont {Campbell}, \citenamefont {Hann}, \citenamefont {Iverson}, \citenamefont {Putterman}, \citenamefont {Bohdanowicz}, \citenamefont {Flammia}, \citenamefont {Keller}, \citenamefont {Refael}, \citenamefont {Preskill}, \citenamefont {Jiang}, \citenamefont {Safavi-Naeini}, \citenamefont {Painter},\ and\ \citenamefont {Brand\~ao}}]{chamberland2022ftqc}%
  \BibitemOpen
  \bibfield  {author} {\bibinfo {author} {\bibfnamefont {C.}~\bibnamefont {Chamberland}}, \bibinfo {author} {\bibfnamefont {K.}~\bibnamefont {Noh}}, \bibinfo {author} {\bibfnamefont {P.}~\bibnamefont {Arrangoiz-Arriola}}, \bibinfo {author} {\bibfnamefont {E.~T.}\ \bibnamefont {Campbell}}, \bibinfo {author} {\bibfnamefont {C.~T.}\ \bibnamefont {Hann}}, \bibinfo {author} {\bibfnamefont {J.}~\bibnamefont {Iverson}}, \bibinfo {author} {\bibfnamefont {H.}~\bibnamefont {Putterman}}, \bibinfo {author} {\bibfnamefont {T.~C.}\ \bibnamefont {Bohdanowicz}}, \bibinfo {author} {\bibfnamefont {S.~T.}\ \bibnamefont {Flammia}}, \bibinfo {author} {\bibfnamefont {A.}~\bibnamefont {Keller}}, \bibinfo {author} {\bibfnamefont {G.}~\bibnamefont {Refael}}, \bibinfo {author} {\bibfnamefont {J.}~\bibnamefont {Preskill}}, \bibinfo {author} {\bibfnamefont {L.}~\bibnamefont {Jiang}}, \bibinfo {author} {\bibfnamefont {A.~H.}\ \bibnamefont {Safavi-Naeini}}, \bibinfo {author} {\bibfnamefont {O.}~\bibnamefont {Painter}},\ and\ \bibinfo
  {author} {\bibfnamefont {F.~G.}\ \bibnamefont {Brand\~ao}},\ }\bibfield  {title} {\bibinfo {title} {Building a fault-tolerant quantum computer using concatenated cat codes},\ }\href {https://doi.org/10.1103/PRXQuantum.3.010329} {\bibfield  {journal} {\bibinfo  {journal} {PRX Quantum}\ }\textbf {\bibinfo {volume} {3}},\ \bibinfo {pages} {010329} (\bibinfo {year} {2022})}\BibitemShut {NoStop}%
\bibitem [{\citenamefont {Noh}\ \emph {et~al.}(2020)\citenamefont {Noh}, \citenamefont {Girvin},\ and\ \citenamefont {Jiang}}]{noh2020encoding}%
  \BibitemOpen
  \bibfield  {author} {\bibinfo {author} {\bibfnamefont {K.}~\bibnamefont {Noh}}, \bibinfo {author} {\bibfnamefont {S.~M.}\ \bibnamefont {Girvin}},\ and\ \bibinfo {author} {\bibfnamefont {L.}~\bibnamefont {Jiang}},\ }\bibfield  {title} {\bibinfo {title} {Encoding an oscillator into many oscillators},\ }\href {https://doi.org/10.1103/PhysRevLett.125.080503} {\bibfield  {journal} {\bibinfo  {journal} {Phys. Rev. Lett.}\ }\textbf {\bibinfo {volume} {125}},\ \bibinfo {pages} {080503} (\bibinfo {year} {2020})}\BibitemShut {NoStop}%
\bibitem [{\citenamefont {Guo}\ \emph {et~al.}(2026)\citenamefont {Guo}, \citenamefont {Mueller},\ and\ \citenamefont {Liu}}]{guo2025concatenateddualdisplacementcode}%
  \BibitemOpen
  \bibfield  {author} {\bibinfo {author} {\bibfnamefont {F.}~\bibnamefont {Guo}}, \bibinfo {author} {\bibfnamefont {F.}~\bibnamefont {Mueller}},\ and\ \bibinfo {author} {\bibfnamefont {Y.}~\bibnamefont {Liu}},\ }\bibfield  {title} {\bibinfo {title} {Concatenated dual displacement code for continuous-variable quantum error correction},\ }\href {https://doi.org/10.1103/d1zs-cy4t} {\bibfield  {journal} {\bibinfo  {journal} {Phys. Rev. Res.}\ }\textbf {\bibinfo {volume} {8}},\ \bibinfo {pages} {023158} (\bibinfo {year} {2026})}\BibitemShut {NoStop}%
\bibitem [{\citenamefont {Xu}\ \emph {et~al.}(2023)\citenamefont {Xu}, \citenamefont {Wang}, \citenamefont {Kuo},\ and\ \citenamefont {Albert}}]{xu2023qubit-oscillator}%
  \BibitemOpen
  \bibfield  {author} {\bibinfo {author} {\bibfnamefont {Y.}~\bibnamefont {Xu}}, \bibinfo {author} {\bibfnamefont {Y.}~\bibnamefont {Wang}}, \bibinfo {author} {\bibfnamefont {E.-J.}\ \bibnamefont {Kuo}},\ and\ \bibinfo {author} {\bibfnamefont {V.~V.}\ \bibnamefont {Albert}},\ }\bibfield  {title} {\bibinfo {title} {Qubit-oscillator concatenated codes: Decoding formalism and code comparison},\ }\href {https://doi.org/10.1103/PRXQuantum.4.020342} {\bibfield  {journal} {\bibinfo  {journal} {PRX Quantum}\ }\textbf {\bibinfo {volume} {4}},\ \bibinfo {pages} {020342} (\bibinfo {year} {2023})}\BibitemShut {NoStop}%
\bibitem [{\citenamefont {Wu}\ \emph {et~al.}(2023)\citenamefont {Wu}, \citenamefont {Brady},\ and\ \citenamefont {Zhuang}}]{Wu2023optimalencodingof}%
  \BibitemOpen
  \bibfield  {author} {\bibinfo {author} {\bibfnamefont {J.}~\bibnamefont {Wu}}, \bibinfo {author} {\bibfnamefont {A.~J.}\ \bibnamefont {Brady}},\ and\ \bibinfo {author} {\bibfnamefont {Q.}~\bibnamefont {Zhuang}},\ }\bibfield  {title} {\bibinfo {title} {Optimal encoding of oscillators into more oscillators},\ }\href {https://doi.org/10.22331/q-2023-08-16-1082} {\bibfield  {journal} {\bibinfo  {journal} {{Quantum}}\ }\textbf {\bibinfo {volume} {7}},\ \bibinfo {pages} {1082} (\bibinfo {year} {2023})}\BibitemShut {NoStop}%
\bibitem [{\citenamefont {Zhuang}\ \emph {et~al.}(2020)\citenamefont {Zhuang}, \citenamefont {Preskill},\ and\ \citenamefont {Jiang}}]{Zhuang_2020}%
  \BibitemOpen
  \bibfield  {author} {\bibinfo {author} {\bibfnamefont {Q.}~\bibnamefont {Zhuang}}, \bibinfo {author} {\bibfnamefont {J.}~\bibnamefont {Preskill}},\ and\ \bibinfo {author} {\bibfnamefont {L.}~\bibnamefont {Jiang}},\ }\bibfield  {title} {\bibinfo {title} {Distributed quantum sensing enhanced by continuous-variable error correction},\ }\href {https://doi.org/10.1088/1367-2630/ab7257} {\bibfield  {journal} {\bibinfo  {journal} {New Journal of Physics}\ }\textbf {\bibinfo {volume} {22}},\ \bibinfo {pages} {022001} (\bibinfo {year} {2020})}\BibitemShut {NoStop}%
\bibitem [{\citenamefont {Zhou}\ \emph {et~al.}(2022)\citenamefont {Zhou}, \citenamefont {Brady},\ and\ \citenamefont {Zhuang}}]{zhou2022enhanching}%
  \BibitemOpen
  \bibfield  {author} {\bibinfo {author} {\bibfnamefont {B.}~\bibnamefont {Zhou}}, \bibinfo {author} {\bibfnamefont {A.~J.}\ \bibnamefont {Brady}},\ and\ \bibinfo {author} {\bibfnamefont {Q.}~\bibnamefont {Zhuang}},\ }\bibfield  {title} {\bibinfo {title} {Enhancing distributed sensing with imperfect error correction},\ }\href {https://doi.org/10.1103/PhysRevA.106.012404} {\bibfield  {journal} {\bibinfo  {journal} {Phys. Rev. A}\ }\textbf {\bibinfo {volume} {106}},\ \bibinfo {pages} {012404} (\bibinfo {year} {2022})}\BibitemShut {NoStop}%
\bibitem [{\citenamefont {Wu}\ \emph {et~al.}(2022)\citenamefont {Wu}, \citenamefont {Zhang},\ and\ \citenamefont {Zhuang}}]{Wu_2022_QST}%
  \BibitemOpen
  \bibfield  {author} {\bibinfo {author} {\bibfnamefont {B.-H.}\ \bibnamefont {Wu}}, \bibinfo {author} {\bibfnamefont {Z.}~\bibnamefont {Zhang}},\ and\ \bibinfo {author} {\bibfnamefont {Q.}~\bibnamefont {Zhuang}},\ }\bibfield  {title} {\bibinfo {title} {Continuous-variable quantum repeaters based on bosonic error-correction and teleportation: architecture and applications},\ }\href {https://doi.org/10.1088/2058-9565/ac4f6b} {\bibfield  {journal} {\bibinfo  {journal} {Quantum Science and Technology}\ }\textbf {\bibinfo {volume} {7}},\ \bibinfo {pages} {025018} (\bibinfo {year} {2022})}\BibitemShut {NoStop}%
\bibitem [{\citenamefont {Brun}\ \emph {et~al.}(2006)\citenamefont {Brun}, \citenamefont {Devetak},\ and\ \citenamefont {Hsieh}}]{brun2006correcting}%
  \BibitemOpen
  \bibfield  {author} {\bibinfo {author} {\bibfnamefont {T.}~\bibnamefont {Brun}}, \bibinfo {author} {\bibfnamefont {I.}~\bibnamefont {Devetak}},\ and\ \bibinfo {author} {\bibfnamefont {M.-H.}\ \bibnamefont {Hsieh}},\ }\bibfield  {title} {\bibinfo {title} {Correcting quantum errors with entanglement},\ }\href {https://doi.org/10.1126/science.1131563} {\bibfield  {journal} {\bibinfo  {journal} {Science}\ }\textbf {\bibinfo {volume} {314}},\ \bibinfo {pages} {436–439} (\bibinfo {year} {2006})}\BibitemShut {NoStop}%
\bibitem [{\citenamefont {Brun}\ \emph {et~al.}(2014)\citenamefont {Brun}, \citenamefont {Devetak},\ and\ \citenamefont {Hsieh}}]{brun2014catalytic}%
  \BibitemOpen
  \bibfield  {author} {\bibinfo {author} {\bibfnamefont {T.~A.}\ \bibnamefont {Brun}}, \bibinfo {author} {\bibfnamefont {I.}~\bibnamefont {Devetak}},\ and\ \bibinfo {author} {\bibfnamefont {M.-H.}\ \bibnamefont {Hsieh}},\ }\bibfield  {title} {\bibinfo {title} {Catalytic quantum error correction},\ }\href {https://doi.org/10.1109/tit.2014.2313559} {\bibfield  {journal} {\bibinfo  {journal} {IEEE Transactions on Information Theory}\ }\textbf {\bibinfo {volume} {60}},\ \bibinfo {pages} {3073–3089} (\bibinfo {year} {2014})}\BibitemShut {NoStop}%
\bibitem [{\citenamefont {Wilde}\ \emph {et~al.}(2007)\citenamefont {Wilde}, \citenamefont {Krovi},\ and\ \citenamefont {Brun}}]{wilde2007optics}%
  \BibitemOpen
  \bibfield  {author} {\bibinfo {author} {\bibfnamefont {M.~M.}\ \bibnamefont {Wilde}}, \bibinfo {author} {\bibfnamefont {H.}~\bibnamefont {Krovi}},\ and\ \bibinfo {author} {\bibfnamefont {T.~A.}\ \bibnamefont {Brun}},\ }\bibfield  {title} {\bibinfo {title} {Entanglement-assisted quantum error correction with linear optics},\ }\href {https://doi.org/10.1103/PhysRevA.76.052308} {\bibfield  {journal} {\bibinfo  {journal} {Phys. Rev. A}\ }\textbf {\bibinfo {volume} {76}},\ \bibinfo {pages} {052308} (\bibinfo {year} {2007})}\BibitemShut {NoStop}%
\bibitem [{\citenamefont {Menicucci}(2014)}]{menicucci2014fault}%
  \BibitemOpen
  \bibfield  {author} {\bibinfo {author} {\bibfnamefont {N.~C.}\ \bibnamefont {Menicucci}},\ }\bibfield  {title} {\bibinfo {title} {Fault-tolerant measurement-based quantum computing with continuous-variable cluster states},\ }\href {https://doi.org/10.1103/PhysRevLett.112.120504} {\bibfield  {journal} {\bibinfo  {journal} {Phys. Rev. Lett.}\ }\textbf {\bibinfo {volume} {112}},\ \bibinfo {pages} {120504} (\bibinfo {year} {2014})}\BibitemShut {NoStop}%
\bibitem [{\citenamefont {Glancy}\ and\ \citenamefont {Knill}(2006)}]{glancy2006error}%
  \BibitemOpen
  \bibfield  {author} {\bibinfo {author} {\bibfnamefont {S.}~\bibnamefont {Glancy}}\ and\ \bibinfo {author} {\bibfnamefont {E.}~\bibnamefont {Knill}},\ }\bibfield  {title} {\bibinfo {title} {Error analysis for encoding a qubit in an oscillator},\ }\href {https://doi.org/10.1103/PhysRevA.73.012325} {\bibfield  {journal} {\bibinfo  {journal} {Phys. Rev. A}\ }\textbf {\bibinfo {volume} {73}},\ \bibinfo {pages} {012325} (\bibinfo {year} {2006})}\BibitemShut {NoStop}%
\bibitem [{\citenamefont {Schmidt}\ and\ \citenamefont {van Loock}(2022)}]{schmidt2022quantum}%
  \BibitemOpen
  \bibfield  {author} {\bibinfo {author} {\bibfnamefont {F.}~\bibnamefont {Schmidt}}\ and\ \bibinfo {author} {\bibfnamefont {P.}~\bibnamefont {van Loock}},\ }\bibfield  {title} {\bibinfo {title} {Quantum error correction with higher gottesman-kitaev-preskill codes: Minimal measurements and linear optics},\ }\href {https://doi.org/10.1103/PhysRevA.105.042427} {\bibfield  {journal} {\bibinfo  {journal} {Phys. Rev. A}\ }\textbf {\bibinfo {volume} {105}},\ \bibinfo {pages} {042427} (\bibinfo {year} {2022})}\BibitemShut {NoStop}%
\bibitem [{\citenamefont {Knill}(2005)}]{Knill2005scalable}%
  \BibitemOpen
  \bibfield  {author} {\bibinfo {author} {\bibfnamefont {E.}~\bibnamefont {Knill}},\ }\bibfield  {title} {\bibinfo {title} {Scalable quantum computing in the presence of large detected-error rates},\ }\href {https://doi.org/10.1103/PhysRevA.71.042322} {\bibfield  {journal} {\bibinfo  {journal} {Phys. Rev. A}\ }\textbf {\bibinfo {volume} {71}},\ \bibinfo {pages} {042322} (\bibinfo {year} {2005})}\BibitemShut {NoStop}%
\bibitem [{\citenamefont {Walshe}\ \emph {et~al.}(2020)\citenamefont {Walshe}, \citenamefont {Baragiola}, \citenamefont {Alexander},\ and\ \citenamefont {Menicucci}}]{walshe2020continous}%
  \BibitemOpen
  \bibfield  {author} {\bibinfo {author} {\bibfnamefont {B.~W.}\ \bibnamefont {Walshe}}, \bibinfo {author} {\bibfnamefont {B.~Q.}\ \bibnamefont {Baragiola}}, \bibinfo {author} {\bibfnamefont {R.~N.}\ \bibnamefont {Alexander}},\ and\ \bibinfo {author} {\bibfnamefont {N.~C.}\ \bibnamefont {Menicucci}},\ }\bibfield  {title} {\bibinfo {title} {Continuous-variable gate teleportation and bosonic-code error correction},\ }\href {https://doi.org/10.1103/PhysRevA.102.062411} {\bibfield  {journal} {\bibinfo  {journal} {Phys. Rev. A}\ }\textbf {\bibinfo {volume} {102}},\ \bibinfo {pages} {062411} (\bibinfo {year} {2020})}\BibitemShut {NoStop}%
\bibitem [{\citenamefont {Marqversen}\ \emph {et~al.}(2025)\citenamefont {Marqversen}, \citenamefont {Wesenberg}, \citenamefont {Zinner},\ and\ \citenamefont {Andersen}}]{marqversen2025performanceanalysisgkperror}%
  \BibitemOpen
  \bibfield  {author} {\bibinfo {author} {\bibfnamefont {F.~K.}\ \bibnamefont {Marqversen}}, \bibinfo {author} {\bibfnamefont {J.~H.}\ \bibnamefont {Wesenberg}}, \bibinfo {author} {\bibfnamefont {N.~T.}\ \bibnamefont {Zinner}},\ and\ \bibinfo {author} {\bibfnamefont {U.~L.}\ \bibnamefont {Andersen}},\ }\href {https://arxiv.org/abs/2505.14775} {\bibinfo {title} {Performance analysis of gkp error correction}} (\bibinfo {year} {2025}),\ \Eprint {https://arxiv.org/abs/2505.14775} {arXiv:2505.14775 [quant-ph]} \BibitemShut {NoStop}%
\bibitem [{\citenamefont {Lai}\ and\ \citenamefont {Brun}(2013)}]{lai2013entanglement}%
  \BibitemOpen
  \bibfield  {author} {\bibinfo {author} {\bibfnamefont {C.-Y.}\ \bibnamefont {Lai}}\ and\ \bibinfo {author} {\bibfnamefont {T.~A.}\ \bibnamefont {Brun}},\ }\bibfield  {title} {\bibinfo {title} {Entanglement increases the error-correcting ability of quantum error-correcting codes},\ }\href {https://doi.org/10.1103/PhysRevA.88.012320} {\bibfield  {journal} {\bibinfo  {journal} {Phys. Rev. A}\ }\textbf {\bibinfo {volume} {88}},\ \bibinfo {pages} {012320} (\bibinfo {year} {2013})}\BibitemShut {NoStop}%
\bibitem [{\citenamefont {Lai}\ \emph {et~al.}(2014)\citenamefont {Lai}, \citenamefont {Brun},\ and\ \citenamefont {Wilde}}]{lai2014dualities}%
  \BibitemOpen
  \bibfield  {author} {\bibinfo {author} {\bibfnamefont {C.-Y.}\ \bibnamefont {Lai}}, \bibinfo {author} {\bibfnamefont {T.~A.}\ \bibnamefont {Brun}},\ and\ \bibinfo {author} {\bibfnamefont {M.~M.}\ \bibnamefont {Wilde}},\ }\bibfield  {title} {\bibinfo {title} {Dualities and identities for entanglement-assisted quantum codes},\ }\href {https://doi.org/10.1007/s11128-013-0704-8} {\bibfield  {journal} {\bibinfo  {journal} {Quantum information processing}\ }\textbf {\bibinfo {volume} {13}},\ \bibinfo {pages} {957} (\bibinfo {year} {2014})}\BibitemShut {NoStop}%
\bibitem [{\citenamefont {Gaitan}(2008)}]{gaitan2008quantum}%
  \BibitemOpen
  \bibfield  {author} {\bibinfo {author} {\bibfnamefont {F.}~\bibnamefont {Gaitan}},\ }\href {https://books.google.co.in/books?id=zwvlqspyOK8C} {\emph {\bibinfo {title} {Quantum Error Correction and Fault Tolerant Quantum Computing}}}\ (\bibinfo  {publisher} {Taylor \& Francis},\ \bibinfo {year} {2008})\BibitemShut {NoStop}%
\bibitem [{\citenamefont {Dash}\ \emph {et~al.}(2024)\citenamefont {Dash}, \citenamefont {Dutta}, \citenamefont {Srikanth},\ and\ \citenamefont {Banerjee}}]{dash2023concatenating}%
  \BibitemOpen
  \bibfield  {author} {\bibinfo {author} {\bibfnamefont {N.~R.}\ \bibnamefont {Dash}}, \bibinfo {author} {\bibfnamefont {S.}~\bibnamefont {Dutta}}, \bibinfo {author} {\bibfnamefont {R.}~\bibnamefont {Srikanth}},\ and\ \bibinfo {author} {\bibfnamefont {S.}~\bibnamefont {Banerjee}},\ }\bibfield  {title} {\bibinfo {title} {Concatenating quantum error-correcting codes with decoherence-free subspaces and vice versa},\ }\href {https://doi.org/10.1103/PhysRevA.109.062411} {\bibfield  {journal} {\bibinfo  {journal} {Phys. Rev. A}\ }\textbf {\bibinfo {volume} {109}},\ \bibinfo {pages} {062411} (\bibinfo {year} {2024})}\BibitemShut {NoStop}%
\bibitem [{\citenamefont {Lai}\ and\ \citenamefont {Brun}(2012)}]{lai2012entanglement}%
  \BibitemOpen
  \bibfield  {author} {\bibinfo {author} {\bibfnamefont {C.-Y.}\ \bibnamefont {Lai}}\ and\ \bibinfo {author} {\bibfnamefont {T.~A.}\ \bibnamefont {Brun}},\ }\bibfield  {title} {\bibinfo {title} {Entanglement-assisted quantum error-correcting codes with imperfect ebits},\ }\href {https://doi.org/10.1103/PhysRevA.86.032319} {\bibfield  {journal} {\bibinfo  {journal} {Phys. Rev. A}\ }\textbf {\bibinfo {volume} {86}},\ \bibinfo {pages} {032319} (\bibinfo {year} {2012})}\BibitemShut {NoStop}%
\bibitem [{\citenamefont {Cerf}\ \emph {et~al.}(2007)\citenamefont {Cerf}, \citenamefont {Leuchs},\ and\ \citenamefont {Polzik}}]{cerf2007quantum}%
  \BibitemOpen
  \bibfield  {author} {\bibinfo {author} {\bibfnamefont {N.~J.}\ \bibnamefont {Cerf}}, \bibinfo {author} {\bibfnamefont {G.}~\bibnamefont {Leuchs}},\ and\ \bibinfo {author} {\bibfnamefont {E.~S.}\ \bibnamefont {Polzik}},\ }\href@noop {} {\emph {\bibinfo {title} {Quantum information with continuous variables of atoms and light}}}\ (\bibinfo  {publisher} {World Scientific},\ \bibinfo {year} {2007})\BibitemShut {NoStop}%
\bibitem [{\citenamefont {Puri}\ \emph {et~al.}(2019)\citenamefont {Puri}, \citenamefont {Grimm}, \citenamefont {Campagne-Ibarcq}, \citenamefont {Eickbusch}, \citenamefont {Noh}, \citenamefont {Roberts}, \citenamefont {Jiang}, \citenamefont {Mirrahimi}, \citenamefont {Devoret},\ and\ \citenamefont {Girvin}}]{puri2019stabilized}%
  \BibitemOpen
  \bibfield  {author} {\bibinfo {author} {\bibfnamefont {S.}~\bibnamefont {Puri}}, \bibinfo {author} {\bibfnamefont {A.}~\bibnamefont {Grimm}}, \bibinfo {author} {\bibfnamefont {P.}~\bibnamefont {Campagne-Ibarcq}}, \bibinfo {author} {\bibfnamefont {A.}~\bibnamefont {Eickbusch}}, \bibinfo {author} {\bibfnamefont {K.}~\bibnamefont {Noh}}, \bibinfo {author} {\bibfnamefont {G.}~\bibnamefont {Roberts}}, \bibinfo {author} {\bibfnamefont {L.}~\bibnamefont {Jiang}}, \bibinfo {author} {\bibfnamefont {M.}~\bibnamefont {Mirrahimi}}, \bibinfo {author} {\bibfnamefont {M.~H.}\ \bibnamefont {Devoret}},\ and\ \bibinfo {author} {\bibfnamefont {S.~M.}\ \bibnamefont {Girvin}},\ }\bibfield  {title} {\bibinfo {title} {Stabilized cat in a driven nonlinear cavity: A fault-tolerant error syndrome detector},\ }\href {https://doi.org/10.1103/PhysRevX.9.041009} {\bibfield  {journal} {\bibinfo  {journal} {Phys. Rev. X}\ }\textbf {\bibinfo {volume} {9}},\ \bibinfo {pages} {041009} (\bibinfo {year} {2019})}\BibitemShut {NoStop}%
\bibitem [{\citenamefont {Cochrane}\ \emph {et~al.}(1999)\citenamefont {Cochrane}, \citenamefont {Milburn},\ and\ \citenamefont {Munro}}]{cochrane1999cat}%
  \BibitemOpen
  \bibfield  {author} {\bibinfo {author} {\bibfnamefont {P.~T.}\ \bibnamefont {Cochrane}}, \bibinfo {author} {\bibfnamefont {G.~J.}\ \bibnamefont {Milburn}},\ and\ \bibinfo {author} {\bibfnamefont {W.~J.}\ \bibnamefont {Munro}},\ }\bibfield  {title} {\bibinfo {title} {Macroscopically distinct quantum-superposition states as a bosonic code for amplitude damping},\ }\href {https://doi.org/10.1103/PhysRevA.59.2631} {\bibfield  {journal} {\bibinfo  {journal} {Phys. Rev. A}\ }\textbf {\bibinfo {volume} {59}},\ \bibinfo {pages} {2631} (\bibinfo {year} {1999})}\BibitemShut {NoStop}%
\bibitem [{\citenamefont {Hastrup}\ and\ \citenamefont {Andersen}(2022)}]{hastrup2022cat}%
  \BibitemOpen
  \bibfield  {author} {\bibinfo {author} {\bibfnamefont {J.}~\bibnamefont {Hastrup}}\ and\ \bibinfo {author} {\bibfnamefont {U.~L.}\ \bibnamefont {Andersen}},\ }\bibfield  {title} {\bibinfo {title} {All-optical cat-code quantum error correction},\ }\href {https://doi.org/10.1103/PhysRevResearch.4.043065} {\bibfield  {journal} {\bibinfo  {journal} {Phys. Rev. Res.}\ }\textbf {\bibinfo {volume} {4}},\ \bibinfo {pages} {043065} (\bibinfo {year} {2022})}\BibitemShut {NoStop}%
\bibitem [{\citenamefont {Michael}\ \emph {et~al.}(2016)\citenamefont {Michael}, \citenamefont {Silveri}, \citenamefont {Brierley}, \citenamefont {Albert}, \citenamefont {Salmilehto}, \citenamefont {Jiang},\ and\ \citenamefont {Girvin}}]{michael2016binomial}%
  \BibitemOpen
  \bibfield  {author} {\bibinfo {author} {\bibfnamefont {M.~H.}\ \bibnamefont {Michael}}, \bibinfo {author} {\bibfnamefont {M.}~\bibnamefont {Silveri}}, \bibinfo {author} {\bibfnamefont {R.~T.}\ \bibnamefont {Brierley}}, \bibinfo {author} {\bibfnamefont {V.~V.}\ \bibnamefont {Albert}}, \bibinfo {author} {\bibfnamefont {J.}~\bibnamefont {Salmilehto}}, \bibinfo {author} {\bibfnamefont {L.}~\bibnamefont {Jiang}},\ and\ \bibinfo {author} {\bibfnamefont {S.~M.}\ \bibnamefont {Girvin}},\ }\bibfield  {title} {\bibinfo {title} {New class of quantum error-correcting codes for a bosonic mode},\ }\href {https://doi.org/10.1103/PhysRevX.6.031006} {\bibfield  {journal} {\bibinfo  {journal} {Phys. Rev. X}\ }\textbf {\bibinfo {volume} {6}},\ \bibinfo {pages} {031006} (\bibinfo {year} {2016})}\BibitemShut {NoStop}%
\bibitem [{\citenamefont {Laha}\ and\ \citenamefont {van Loock}(2026)}]{laha2026binomial}%
  \BibitemOpen
  \bibfield  {author} {\bibinfo {author} {\bibfnamefont {P.}~\bibnamefont {Laha}}\ and\ \bibinfo {author} {\bibfnamefont {P.}~\bibnamefont {van Loock}},\ }\bibfield  {title} {\bibinfo {title} {Arbitrary high-fidelity binomial codes from multiphoton spin-boson interactions},\ }\href {https://doi.org/10.1103/58jr-l1x6} {\bibfield  {journal} {\bibinfo  {journal} {Phys. Rev. Res.}\ }\textbf {\bibinfo {volume} {8}},\ \bibinfo {pages} {013237} (\bibinfo {year} {2026})}\BibitemShut {NoStop}%
\end{thebibliography}%
\end{document}